\documentclass[aps,prb,twocolumn,longbibliography,showpacs]{revtex4-1}

\usepackage{graphicx}
\usepackage{dcolumn}
\usepackage{bm}
\usepackage[version=3]{mhchem}
\usepackage{color}

\begin{document}

\title{Doping Li-rich cathode material \ce{Li2MnO3}: Interplay between lattice site preference, electronic structure, and delithiation mechanism}
\author{Khang Hoang}
\email{khang.hoang@ndsu.edu}
\affiliation{Department of Physics, North Dakota State University, Fargo, North Dakota 58108, USA}

\date{\today}

\begin{abstract}

We report a detailed first-principles study of doping in \ce{Li2MnO3}, in both the dilute doping limit and heavy doping, using hybrid density-functional calculations. We find that Al, Fe, Mo, and Ru impurities are energetically most favorable when incorporated into \ce{Li2MnO3} at the Mn site, whereas Mg is most favorable when doped at the Li sites. Ni, on the other hand, can be incorporated at the Li site and/or the Mn site, and the distribution of Ni over the lattice sites can be tuned by tuning the materials preparation conditions. There is a strong interplay between the lattice site preference and charge and spin states of the dopant, the electronic structure of the doped material, and the delithiation mechanism. The calculated electronic structure and voltage profile indicate that, in Ni-, Mo-, or Ru-doped \ce{Li2MnO3}, oxidation occurs on the electrochemically active transition-metal ion(s) before it does on oxygen during the delithiation process. The role of the dopants is to provide charge-compensation and bulk electronic conduction mechanisms in the initial stages of delithiation, hence enabling the oxidation of the lattice oxygen in the later stages. This work thus illustrates how the oxygen-oxidation mechanism can be used in combination with the conventional mechanism involving transition-metal cations in design of high-capacity battery cathode materials.

\end{abstract}


\maketitle


\section{Introduction}\label{sec;intro}

\begin{figure}
\centering
\includegraphics[width=8.4cm]{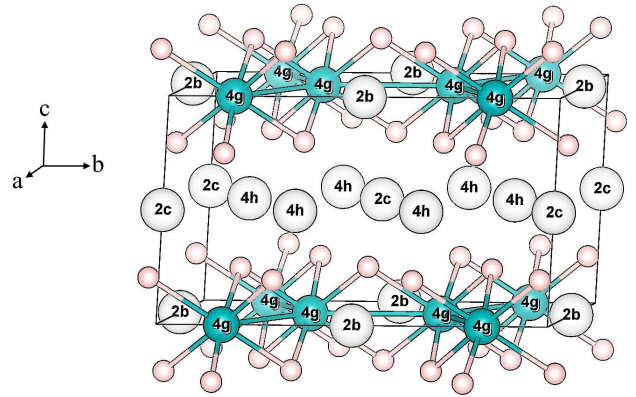}
\vspace{-0.2cm}
\caption{Monoclinic ($C2/m$) structure of Li-rich layered-oxide \ce{Li2MnO3}. Large gray spheres at the $2b$, $2c$, and $4h$ Wyckoff positions are Li, medium blue spheres at the $4g$ Wyckoff position are Mn, and small red spheres are O.}
\label{fig;struct}
\end{figure}

Li-rich layered-oxide \ce{Li2MnO3} and related materials have been of great interest for high-capacity lithium-ion battery electrodes.\cite{Thackeray2005JMC,Croguennec2015JACS} The materials are also found to be intriguing for having an unconventional delithiation mechanism involving anionic redox processes, in contrast or in addition to the conventional mechanism involving redox reactions of transition-metal cations. The title compound, also known as Li[Li$_{1/3}$Mn$_{2/3}$]O$_2$, has a layered structure\cite{Massarotti1997} with Li$^+$ and Mn$^{4+}$ ions occupying the octahedral lattice sites formed by alternating layers of oxygen as shown in Fig.~\ref{fig;struct}. Though initially believed to be electrochemically inactive, it was later discovered that \ce{Li2MnO3} can be made active by, for instance, charging to a high voltage.\cite{Kalyani1999} The undoped compound, however, exhibits only a very limited electrochemical capacity.\cite{Sathiya2013} 

Recently,\cite{Hoang2015PRA} we showed that the intrinsic delithiation mechanism in \ce{Li2MnO3} involves oxidation at the oxygen lattice site where O$^{2-}$ is oxidized to O$^-$ upon lithium extraction. The localized electron hole states on oxygen\cite{Sathiya2013,Luo2016NC} and the associated local lattice distortion constitute an oxygen hole polaron $\eta_{\rm O}^+$ bound to other defects such as the negatively charged lithium vacancy $V_{\rm Li}^-$ (i.e., the void in the lattice left by the removal of a Li$^{+}$ ion); the species was, therefore, referred to as ``bound oxygen hole polaron'' or ``O$^-$ bound polaron.'' The difficulty in activating \ce{Li2MnO3} and the poor electrochemical performance could be attributed in part to the lack of percolation pathways for efficient $\eta_{\rm O}^+$ transport in the bulk at the onset of delithiation where the lithium vacancy concentration is low. We also argued that the electrochemical performance could be improved via ion substitution according to which partial replacement of, e.g., Mn by electrochemically active ions would introduce additional charge-compensation and electronic conduction mechanisms highly needed at the beginning of delithiation.\cite{Hoang2015PRA}

Experimental studies of doping/co-doping in \ce{Li2MnO3} have been carried out by various research groups.\cite{Yu2016JMCA,Xiang2017Ionics,Yuge2017JPS,Luo2016JACS,Tang2017JPCL,Lu2001,Matsunaga2016JPCL,Ma2014CEJ,Mori2011JPS,Sathiya2013} Back in the early 2000s, for example, Lu and Dahn\cite{Lu2001} already considered Li[Li$_{1/3-2x/3}$Mn$_{2/3-x/3}$Ni$_x$]O$_2$, which is a Ni-doped \ce{Li2MnO3} system. Recent investigations of doped \ce{Li2MnO3} materials focused more on understanding the electrochemical activity associated with oxygen redox processes.\cite{Luo2016JACS,Sathiya2013} On the theory side, a number of computational studies have been carried out for \ce{Li2MnO3} doped with various impurities,\cite{Kong2015JMCA,Kong2015JPCC,Lee2015JPS,Gao2014JMCA,Gao2015CM,Yang2017AFM} using density-functional theory (DFT) with the on-site Hubbard correction (DFT$+U$).\cite{Anisimov1997JPCM} Yet, a detailed understanding of the lattice site preference of the dopants in the material under different preparation conditions as well as effects of doping on the electronic structure and hence the delithiation mechanism is still lacking. Besides, DFT$+U$ with $U$ applied only on the transition-metal $d$ states, used in the previous studies,\cite{Gao2014JMCA,Kong2015JMCA,Kong2015JPCC,Gao2015CM,Lee2015JPS,Yang2017AFM,Koyama2009,Xiao2012,Xiao2012JPCC,Marusczyk2017JMCA} cannot capture the intriguing physics of \ce{Li2MnO3} as discussed in Ref.~\citenum{Hoang2015PRA}, suggesting that a more rigorous approach is required. 

We herein present a detailed study of \ce{Li2MnO3} doped with representative non-transition-metal (Mg and Al) and transition-metal (Fe, Ni, Mo, and Ru) impurities, using first-principles calculations based on a hybrid DFT/Hartree-Fock approach in which all electronic states in the materials are treated on equal footing. We start with defect calculations which mainly consider the impurities in the dilute doping limit where the lattice site preference and charge and spin states of the dopants as well as charge-compensation mechanisms in the materials are determined. We then develop structural models to investigate heavily doped \ce{Li2MnO3} systems where the focus is on the electronic structure and the interplay with the delithiation mechanism. Finally, we present voltage profiles calculated for the doped compounds and discuss the role of doping in \ce{Li2MnO3}-based cathode materials.

\section{Methodology}\label{sec;method}

The calculations are based on DFT, using the Heyd-Scuseria-Ernzerhof (HSE06) functional,\cite{heyd:8207} the projector augmented wave method,\cite{PAW1} and a plane-wave basis set, as implemented in the Vienna {\it Ab Initio} Simulation Package (\textsc{vasp}).\cite{VASP2} The Hartree-Fock mixing parameter ($\alpha$) and the screening length are set to the standard values of 25\% and 10 {\AA}, respectively. It was previously shown that, with these parameters, HSE06 gives the \ce{O2} molecule a binding energy of 5.16 eV, in excellent agreement with experiments (5.12 eV).\cite{Hoang2014JMCA} The ferromagnetic spin configuration is assumed for the transition-metal sublattice in \ce{Li2MnO3} as it was found that the ferromagnetic and antiferromagnetic configurations are almost degenerate in energy.\cite{Hoang2015PRA} The plane-wave basis-set cutoff is set to 500 eV and spin polarization is included. Convergence with respect to self-consistent iterations is assumed when the total energy difference between cycles is less than $10^{-4}$ eV and the residual forces are less than $0.01$ eV/{\AA}.

\begin{figure}
\centering
\includegraphics[width=8.4cm]{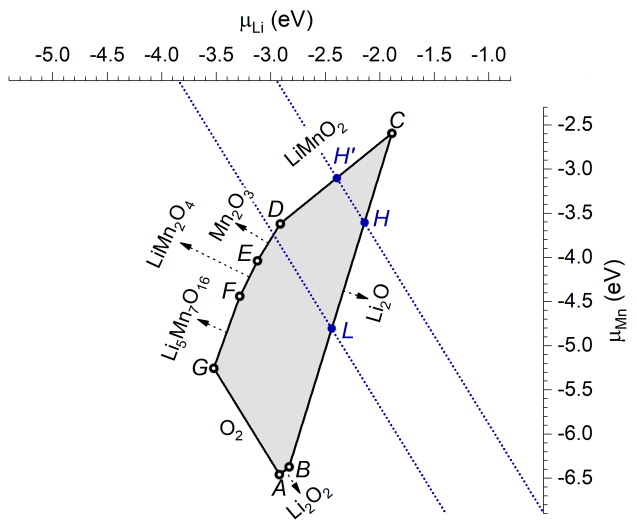}
\vspace{-0.2cm}
\caption{Chemical-potential diagram for \ce{Li2MnO3}, constructed using data from Ref.~\citenum{Hoang2015PRA}. Only Li$-$Mn$-$O phases that define the stability region, shown as a shaded polygon, are indicated. The dotted blue lines correspond to the $\mu_{\rm O}$ levels associated oxygen in air at 500 $^\circ$C and 950 $^\circ$C. Actual synthesis conditions are expected to be in the shaded region enclosed approximately by points $L$, $H$, $H'$, and $D$.}
\label{fig;chempot}
\end{figure}

The impurities in \ce{Li2MnO3} in the dilute doping limit are modeled using 108-atom hexagonal supercells;\cite{Hoang2015PRA} integrations over the supercell Brillouin zone are carried out using the $\Gamma$ point. In these defect calculations, the lattice parameters are fixed to the calculated bulk values but all the internal coordinates are fully relaxed. The formation energy of an intrinsic defect, impurity (dopant), or defect complex X (hereafter often referred commonly to as ``defect'') in charge state $q$ is defined as
\begin{equation}\label{eq;eform}
E^f({\mathrm{X}}^q)=E_{\mathrm{tot}}({\mathrm{X}}^q)-E_{\mathrm{tot}}({\mathrm{bulk}})-\sum_{i}{n_i\mu_i}+q(E_{\mathrm{v}}+\mu_{e})+ \Delta^q ,
\end{equation}
where $E_{\mathrm{tot}}(\mathrm{X}^{q})$ and $E_{\mathrm{tot}}(\mathrm{bulk})$ are, respectively, the total energies of a supercell containing the defect and that of the perfect bulk material. $\mu_{i}$ is the atomic chemical potential of species $i$ (and is referenced to bulk metals or O$_{2}$ molecules at 0 K). $n_{i}$ is the number of atoms of species $i$ that have been added ($n_{i}>$0) or removed ($n_{i}<$0) to form the defect. $\mu_{e}$ is the electronic chemical potential, i.e., the Fermi level, that is, as a convention, referenced to the valence-band maximum (VBM) in the bulk ($E_{\mathrm{v}}$). $\Delta^q$ is the correction term to align the electrostatic potentials of the bulk and defect supercells and to account for finite-size effects on the total energies of charged defects.\cite{Freysoldt2009} In the calculation of $\Delta^q$, a total static dielectric constant of 17.69, previously calculated for \ce{Li2MnO3},\cite{Hoang2015PRA} is used. 

Under thermal equilibrium, the concentration of a defect is related to its formation energy:\cite{walle:3851} 
\begin{equation}\label{eq;con} 
c=N_{\mathrm{sites}}N_{\mathrm{config}}\mathrm{exp}\left(\frac{-E^{f}}{k_{B}T}\right), 
\end{equation} 
where $N_{\mathrm{sites}}$ is the number of high-symmetry sites in the lattice per unit volume on which the defect can be incorporated, $N_{\mathrm{config}}$ is the number of equivalent configurations (per site), and $k_{B}$ is Boltzmann's constant. Equation (\ref{eq;con}) indicates that $c$ is higher at higher temperatures and/or when the defect has a lower formation energy.

The atomic chemical potentials of Li, Mn, and O in \ce{Li2MnO3} are not arbitrary but subject to thermodynamic constraints which can be used to represent actual experimental conditions, e.g., during materials preparation. Figure~\ref{fig;chempot} shows the chemical-potential diagram for \ce{Li2MnO3}, produced with data from Ref.~\citenum{Hoang2015PRA}. The region in which the compound is thermodynamically stable is determined by considering equilibria with other Li$-$Mn$-$O phases.\cite{Hoang2015PRA} In principle, $\mu_{\rm Li}$, $\mu_{\rm Mn}$, and $\mu_{\rm O}$ can take any values within the stability region in which
\begin{equation}\label{eq;li2mno3} 
2\mu_{\rm Li}+\mu_{\rm Mn}+3\mu_{\rm O}=\Delta H^{f}({\rm Li}_{2}{\rm Mn}{\rm O}_{3}),
\end{equation}
where $\Delta H^{f}$ is the formation enthalpy (previously calculated to be $-12.30$ eV per formula unit,\cite{Hoang2015PRA} in good agreement with the experimental value $12.78\pm 0.05$ eV\cite{Cupid2016JCSJ} reported more recently). More realistic $\mu_i$ values, however, can be identified by taking into account actual experimental conditions. For example, \ce{Li2MnO3} is often prepared by solid-state reaction in the temperature range from 500$^\circ$C to 950$^\circ$C.\cite{Massarotti1997,Kalyani1999,Robertson2002,Kubota2012} The synthesis conditions can thus be identified with the shaded region between the two dotted lines in Fig.~\ref{fig;chempot} which corresponds to the range of the oxygen chemical potential $\mu_{\rm O} = -0.87$ eV (e.g., at point $L$) to $-1.47$ eV (points $H$ and $H'$).\cite{Hoang2015PRA} This region is enclosed approximately by points $L$, $H$, $H'$, and $D$. Note that $\mu_{\rm O}$ is related to temperature and the oxygen partial pressure; it is lower at higher temperatures and/or lower oxygen partial pressures and/or in the presence of oxygen-reducing agents.\cite{Hoang2015PRA} In Fig.~\ref{fig;chempot}, $\mu_{\rm O}=0$ eV along the \ce{O2} line (which also corresponds to $T=0$ K).

For the impurities in \ce{Li2MnO3}, their chemical potential values are between minus infinity and zero (with respect to the total energy per atom of the bulk metals). More realistic values can be estimated based on solubility-limiting phases that may be formed between the impurities and the constituting elements of the host compound.\cite{walle:3851} In the following, the chemical potentials of Mg, Al, Fe, Ni, Mo, and Ru impurities are set as $\mu_{\rm Mg} = -7.0$ eV, $\mu_{\rm Al} = -9.0$ eV, $\mu_{\rm Fe} = -5.0$ eV, $\mu_{\rm Ni} = -4.0$ eV, $\mu_{\rm Mo} = -7.0$ eV, and $\mu_{\rm Ru} = -5.0$ eV. These chemical potential values are chosen somewhat arbitrarily; however, the choice does not affect the physics of what we are presenting as we are interested only in the relative formation energies of the impurities.

Heavy doping of \ce{Li2MnO3} with Ni, Mo, or Ru is simulated using smaller, monoclinic cells containing 24 atoms. Starting with the conventional unit cell of undoped \ce{Li2MnO3} as shown in Fig.~\ref{fig;struct}, Mn-site doping corresponds to replacing one of the Mn atoms at the $4g$ lattice site with the dopant. As for Li-site doping, the dopant replaces one of the Li atoms either at the $2b$, $2c$, or $4h$ lattice site. Simultaneous doping at the Li site and the Mn site is also considered. Voltage profiles are calculated using the 24-atom cell size and the expression for the deintercalation voltage derived previously,\cite{Aydinol1997,Hoang2015PRA} assuming topotactic transitions between compositions with different lithium concentrations. In these calculations, integrations over the Brillouin zone are carried out using a $\Gamma$-centered $3\times 2 \times 3$ $k$-point mesh; $6\times 3 \times 6$ or denser $k$-point meshes are used in calculations to obtain high-quality electronic densities of states. Additional calculations make use of monoclinic supercells containing 48 atoms. Note that, unlike in the dilute doping case, the cell volume and shape and internal coordinates are all relaxed in the calculations for heavily doped \ce{Li2MnO3}.

Finally, we note that calculations of bulk \ce{Li2MnO3} using the HSE06 functional but with a smaller Hartree-Fock mixing parameter, e.g., $\alpha=17\%$, have also been reported in the literature.\cite{Seo2016NC} In our calculations, we find both the standard $\alpha$ value and $\alpha =17\%$ give Li$_{2-x}$MnO$_3$ ($0\leq x\leq2$) similar electronic structures, especially the nature of the electronic states near the band edges; the difference is mainly in the calculated band gaps, see Ref.~\citenum{SM}. The results also indicate that \ce{Li2MnO3} remains non-mettalic upon lithium extraction.\cite{SM} Also note that previous calculations using DFT$+U$ with the $U$ term applied only on the Mn $3d$ orbitals show that \ce{Li2MnO3} becomes a metal upon lithium extraction,\cite{Koyama2009,Xiao2012,Xiao2012JPCC,Yang2017AFM,Marusczyk2017JMCA} which is in contrast to the results obtained using the HSE06 functional\cite{Hoang2015PRA,Chen2016CM,SM} or DFT$+U$ with the on-site Hubbard corrections applied on both the Mn $3d$ and O $2p$ states\cite{Hoang2015PRA} according to which the material remains non-metallic. Experiments should, therefore, be carried out to determine the nature of bulk electronic conduction in partially delithiated Li$_{2-x}$MnO$_3$ (especially in the nearly fully lithiated regime, i.e., $x\ll 1$); the results may also be able to validate one computational method over another.

\section{Results and Discussion}\label{results}

\subsection{Lattice site preference and charge and spin states in the dilute doping limit}\label{sec;dilute}

\begin{figure}
\centering
\includegraphics[width=8.6cm]{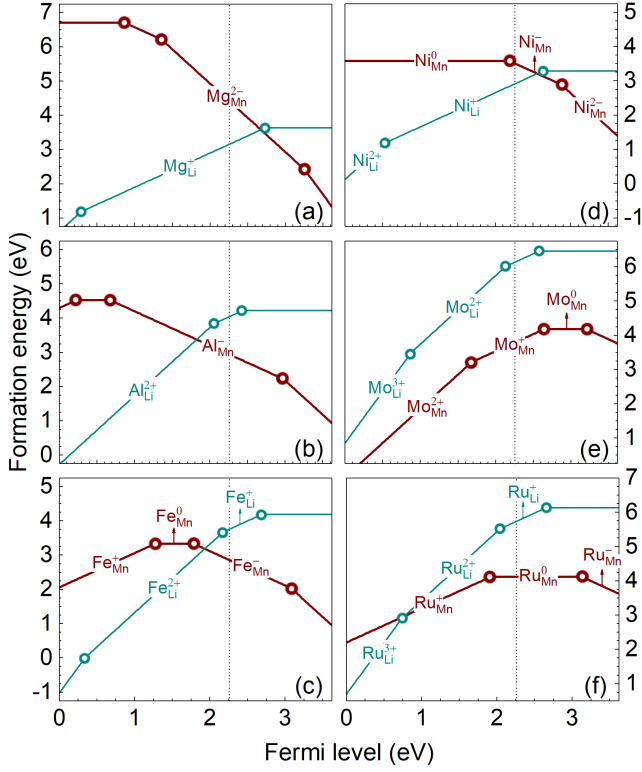}
\vspace{-0.2cm}
\caption{Formation energies of substitutional impurities at the Li site or the Mn site in \ce{Li2MnO3} obtained at point $H$ in the chemical-potential diagram (Fig.~\ref{fig;chempot}) and plotted as a function of Fermi level from the VBM to the conduction-band minimum (CBM): (a) Mg, (b) Al, (c) Fe, (d) Ni, (e) Mo, and (f) Ru. The slope in the energy plots indicates charge state $q$: Positively (negatively) charged defects have positive (negative) slopes. For each defect, only true charge states are indicated. The vertical dotted line marks the Fermi level of undoped \ce{Li2MnO3}, $\mu_{e}^{\rm{int}}$, determined by the intrinsic defects.\cite{Hoang2015PRA}} 
\label{fig;fe}
\end{figure}

Figure~\ref{fig;fe} shows the formation energies of Mg, Al, Fe, Ni, Mo, and Ru incorporated into \ce{Li2MnO3} at the Li site or the Mn site, obtained under the conditions at point $H$ in the chemical-potential diagram. Calculations are carried out for the impurities at all three inequivalent Li lattice sites, however, only the lowest-energy configuration is reported. Among possible values of $q$, true charge states are indicated in the figure and their corresponding defect configurations are called elementary defects; the other (nominal) charge states correspond to complexes consisting of the elementary defects and electron polaron(s) $\eta^-$ or electron-hole-related species. As reported in Ref.~\citenum{Hoang2015PRA}, Mn in undoped \ce{Li2MnO3} is stable as Mn$^{4+}$ and $\eta^-$ is the localized electron and local lattice distortion associated with the high-spin Mn$^{3+}$ ion at the Mn site; the hole-related species can be oxygen hole polaron $\eta_{\rm O}^+$ (i.e., a hole localized on an O ion or, equivalently, O$^-$) or, occasionally, $\eta_{\rm O\ast}^+$ (a hole delocalized over two O ions), depending on specific defect complexes and their effective charge states. $\eta_{\rm O}^+$ is a bound polaron; i.e., it is stable only in the presence of a negatively charged defect such as lithium vacancy $V_{\rm Li}^-$.\cite{Hoang2015PRA} Mg$_{\rm Li}^0$, for example, is a complex of Mg$_{\rm Li}^+$ (i.e., Mg$^{2+}$ at the Li site) and $\eta^-$; Mg$_{\rm Mn}^-$ is a complex of Mg$_{\rm Mn}^{2-}$ (i.e., Mg$^{2+}$ at the Mn site) and $\eta_{\rm O}^+$; see Fig.~\ref{fig;fe}(a). $\eta_{\rm O\ast}^+$ is found to be present only in Al$_{\rm Mn}^0$, a complex of Al$_{\rm Mn}^-$ (i.e., Al$^{3+}$ at the Mn site) and $\eta_{\rm O\ast}^+$, and in Al$_{\rm Mn}^+$, a complex of Al$_{\rm Mn}^-$, $\eta_{\rm O}^+$, and $\eta_{\rm O\ast}^+$. 

\begin{figure}
\centering
\includegraphics[width=8.6cm]{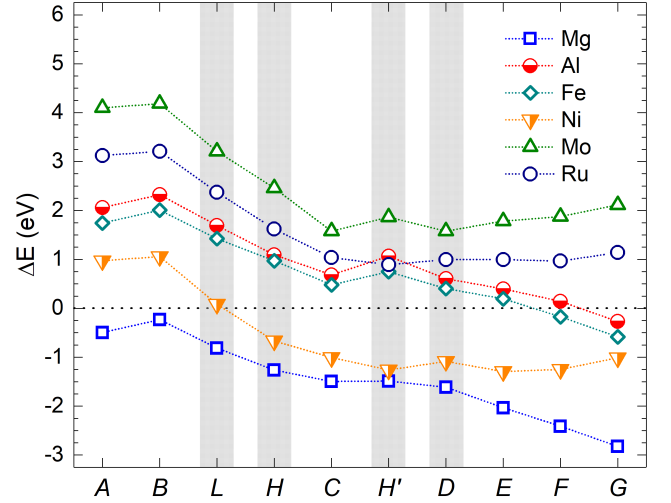}
\vspace{-0.2cm}
\caption{Difference between the formation energy at the Li site and that at the Mn site, obtained under the conditions at different points in the chemical-potential diagram (Fig.~\ref{fig;chempot}). $\Delta E > 0$ means the impurity is energetically more favorable at the Mn lattice site. Points $L$, $H$, $H'$, and $D$ can be regarded as being close to actual conditions during materials preparation.}
\label{fig;fe;diff}
\end{figure}

The lattice site preference is determined by examining the formation energy of the energetically most stable configuration of the impurities at the Li site and that at the Mn site.\cite{Hoang2012JPS,Hoang2017PRMoxides} For a given set of the atomic chemical potentials, which corresponds to a point in the chemical-potential diagram, the energies are obtained at the Fermi level of undoped \ce{Li2MnO3} ($\mu_{e}^{\rm{int}}$, indicated by the vertical dotted line in Fig.~\ref{fig;fe}), determined by intrinsic defects.\cite{Hoang2015PRA} Figure~\ref{fig;fe} indicates that, under the conditions at point $H$, e.g., Mg is energetically most stable as Mg$_{\rm Li}^+$; Al as Al$_{\rm Mn}^-$; Fe as Fe$_{\rm Mn}^-$ (high-spin Fe$^{3+}$ at the Mn site); Ni as Ni$_{\rm Li}^+$ (Ni$^{2+}$ at the Li site); Mo as Mo$_{\rm Mn}^+$ (Mo$^{5+}$ at the Mn site); and Ru as Ru$_{\rm Mn}^0$ (low-spin Ru$^{4+}$ at the Mn site). 

We note that the energies of Mg$_{\rm Li}^+$, as an isolated defect, at the $2c$ site and the $4h$ site are degenerate and lower than the energy at the $2b$ site by 0.04 eV. Isolated Ni$_{\rm Li}^+$, on the other hand, has the lowest energy at the $2b$ lattice site; the energy at the $2c$ site or the $4h$ site is higher by 0.08 eV. These energy differences are independent of the choice of atomic chemical potentials.

Figure~\ref{fig;fe;diff} shows the formation-energy difference, $\Delta E$, between the Li site and the Mn site under the conditions at different points in the chemical-potential diagram. $\Delta E$ is defined as the difference between the formation energy (at $\mu_{e}^{\rm{int}}$) of the lowest-energy defect configuration at the Li site and that at the Mn site.\cite{Hoang2017PRMoxides} $\Delta E > 0$ means the impurity is energetically more favorable at the Mn site than at the Li site; for $\Delta E \sim 0$, one should expect that the impurity can be incorporated simultaneously at both the lattice sites with almost equal concentrations. The results show that the lattice site preference of the impurities can be dependent on the relative abundance of the host constituents during synthesis, expressed in terms of the atomic chemical potentials in our computational approach. Focusing on the conditions at points $L$, $H$, $H'$, and $D$, we find that Al, Fe, Mo, and Ru would be incorporated into \ce{Li2MnO3} at the Mn site, whereas Mg would be at the Li site. These results are thus in agreement with the corresponding doped \ce{Li2MnO3} compositions observed in experiments.\cite{Yu2016JMCA,Xiang2017Ionics,Yuge2017JPS,Ma2014CEJ,Mori2011JPS} Yu et al.,\cite{Yu2016JMCA} for example, reported that Mg occupies the Li ($4h$) site in \ce{Li2MnO3}, which is consistent with the results for Mg. We note that, though Mg$_{\rm Li}^+$ has the same energy at the $4h$ site and the $2c$ site as mentioned above, its concentration at $4h$ is higher than that at $2c$ because the number of the $4h$ sites in the supercell is twice that of the $2c$ sites.        

Interestingly, under the conditions at points $H$, $H'$, and $D$, Ni is energetically most favorable when incorporated at the Li ($2b$) site, whereas, at point $L$, the impurity is more favorable at the Mn ($4g$) site and stable as Ni$_{\rm Mn}^-$ (low-spin Ni$^{3+}$ at the Mn site), when considered as an isolated defect. Our results thus indicate that the lattice site preference and hence the ratio between the concentration of Ni$_{\rm Li}$ and that of Ni$_{\rm Mn}$ are dependent on the preparation conditions: a higher Ni$_{\rm Li}$/Ni$_{\rm Mn}$ ratio, for example, can be obtained at higher synthesis temperatures (i.e., lower $\mu_{\rm O}$ values). It is important to keep in mind that, in addition to affecting the formation energy through the chemical potential terms in Eq.~(\ref{eq;eform}), the synthesis temperature also strongly affects the defect concentration via $k_{B}T$ in the exponential in Eq.~(\ref{eq;con}).

The results for Ni are consistent with experiments reported by Matsunaga et al.\cite{Matsunaga2016JPCL} in which Ni was found to be incorporated mainly at the $2b$ site in Li$_2$(Ni$_z$Mn$_{1-z}$)O$_3$ samples annealed at 900 $^\circ$C in air (see Table 1 of Ref.~\citenum{Matsunaga2016JPCL}), though it is quite surprising that no Li was found in the Li layers,\cite{Matsunaga2016JPCL} given that Ni$_{\rm Li}^+$ has a relatively small energy difference between the $2b$ site and the $4h$ ($2c$) site. The reason could be kinetically related and/or due to defect-defect interaction. We note that the experimental conditions used in Ref.~\citenum{Matsunaga2016JPCL} can be regarded as being close to those associated with the $H-H'$ segment in Fig.~\ref{fig;chempot}. 

Let us now briefly compare some of the above results for \ce{Li2MnO3} with those for the same impurities but in layered \ce{LiMO2} (M = Co, Ni, Mn) reported recently.\cite{Hoang2017PRMoxides} It was found that Al and Fe impurities in \ce{LiMO2} are most favorable at the M site, under realistic conditions, similar to the results for Al and Fe in \ce{Li2MnO3}. Regarding Mg, it can be incorporated into \ce{LiMO2} (M = Co, Ni) at the M site or both the M and Li sites; in \ce{LiMnO2}, Mg can be incorporated at the Mn or Li site or both the lattice sites.\cite{Hoang2017PRMoxides} The results are thus different from those for \ce{Li2MnO3} which show that Mg is always more favorable energetically at the Li site; see Fig.~\ref{fig;fe;diff}. As for Ni, the impurity can be incorporated into \ce{LiCoO2} at the Co site and \ce{LiMnO2} at the Mn site or both the Mn and Li sites.\cite{Hoang2017PRMoxides} In \ce{Li2MnO3}, on the other hand, Ni can be incorporated at the Mn site or the Li site or both the lattice sites, as reported earlier. All these results suggest that, in materials such as $z$\ce{Li2MnO3}$\cdot$($1-z$)\ce{LiMO2} composites, the lattice site preference of a dopant may not be the same in \ce{LiMO2}-rich vs. \ce{Li2MnO3}-rich regions.

\begin{table}
\caption{\ Defect models for the impurities in the dilute doping limit, obtained under the conditions at points $L$, $H$, $H'$, and $D$ in the chemical-potential diagram. Only the most stable configurations are reported; other configurations that are close in energy are listed in the footnotes. It is noted that defect-defect interaction may change the relative stability of the complexes and the charge and spin states of the transition-metal impurities; see the text for further discussions.}\label{tab;models}
\begin{center}
\begin{ruledtabular}
\begin{tabular}{lcccc}
&$L$&$H$&$H'$&$D$ \\
\hline
Mg&Mg$_{\rm Li}^+$$-$$\eta^-$&Mg$_{\rm Li}^+$$-$$\eta^-$&Mg$_{\rm Li}^+$$-$$\eta^-$&Mg$_{\rm Li}^+$$-$$V_{\rm Li}^-$$^a$ \\
Al&Al$_{\rm Mn}^-$$-$Li$_i^+$&Al$_{\rm Mn}^-$$-$Li$_i^+$&Al$_{\rm Mn}^-$$-$Mn$_{\rm Li}^+$&Al$_{\rm Mn}^-$$-$Mn$_{\rm Li}^+$ \\
Fe&Fe$_{\rm Mn}^0$&Fe$_{\rm Mn}^0$$^b$&Fe$_{\rm Mn}^-$$-$Mn$_{\rm Li}^+$&Fe$_{\rm Mn}^-$$-$Mn$_{\rm Li}^+$ \\
Ni&Ni$_{\rm Mn}^0$&Ni$_{\rm Li}^+$$-$$\eta^-$&Ni$_{\rm Li}^+$$-$$\eta^-$&Ni$_{\rm Li}^+$$-$$\eta^-$$^c$ \\
Mo&Mo$_{\rm Mn}^0$&Mo$_{\rm Mn}^0$&Mo$_{\rm Mn}^0$&Mo$_{\rm Mn}^+$$-$$V_{\rm Li}^-$$^d$ \\
Ru&Ru$_{\rm Mn}^0$&Ru$_{\rm Mn}^0$&Ru$_{\rm Mn}^0$&Ru$_{\rm Mn}^0$ \\
\end{tabular}
\end{ruledtabular}
\end{center}
\begin{flushleft}
$^a$Mg$_{\rm Li}^+$$-$$\eta^-$ ($+0.00$ eV). $^b$Fe$_{\rm Mn}^-$$-$Li$_i^+$ ($+0.10$ eV). $^c$Ni$_{\rm Li}^+$$-$$V_{\rm Li}^-$ ($+0.16$ eV). $^d$Mo$_{\rm Mn}^0$ ($+0.02$ eV).
\end{flushleft}  
\end{table}

Since many of the stable configurations for the impurities in \ce{Li2MnO3} are positively or negatively charged, charge compensation is needed when they occur in the material, which can be provided by intrinsic point defects or by having the transition-metal impurity ion change its charge state. Explicit calculations for possible low-energy defect complexes between the charged impurities and intrinsic defects based on the results for the latter reported in Ref.~\citenum{Hoang2015PRA} are carried out. Table \ref{tab;models} summarizes the results obtained under the conditions at points $L$, $H$, $H'$, and $D$ in the chemical-potential diagram. We find that Mg$_{\rm Li}^+$ can occur in the form of Mg$_{\rm Li}^+$$-$$\eta^-$; i.e., the incorporation of Mg at the Li site is charge-compensated by the formation of $\eta^-$ (i.e., Mn$^{3+}$) in the host. The complex has a binding energy $E_b = 0.44$ eV with respect to its isolated constituents. At point $D$, this complex becomes degenerate in energy with Mg$_{\rm Li}^+$$-$$V_{\rm Li}^-$ ($E_b = 0.37$ eV). Al$_{\rm Mn}^-$ is most stable in the form of Al$_{\rm Mn}^-$$-$Li$_i^+$ ($E_b = 0.31$ eV) at points $L$ and $H$, where Li$_i^+$ corresponds to an interstitial Li$^+$ ion in the Li layer,\cite{Hoang2015PRA} or Al$_{\rm Mn}^-$$-$Mn$_{\rm Li}^+$ ($E_b = 0.39$ eV) at points $H'$ and $D$, where Mn$_{\rm Li}^+$ corresponds to Mn$^{2+}$ in the Li layer. Fe$_{\rm Mn}^-$, as isolated defect, can occur in the form of Fe$_{\rm Mn}^0$ (i.e., Fe$^{4+}$ at the Mn site); in this case, Fe changes its charge state from $+3$ to $+4$ to maintain charge neutrality. In the presence of other positively charged impurities, however, such as in Li$_{1.23}$Fe$_{0.15}$Ni$_{0.15}$Mn$_{0.46}$O$_2$,\cite{Yuge2017JPS} where Ni is likely stable as Ni$_{\rm Li}^+$, Fe remains as Fe$_{\rm Mn}^-$. 

As for the other impurities, Ni$_{\rm Mn}^+$, as an isolated defect, can occur in the form of Ni$_{\rm Mn}^0$ (i.e., Ni$^{4+}$ at the Mn site) at point $L$; Ni$_{\rm Li}^+$ as Ni$_{\rm Li}^+$$-$$\eta^-$ ($E_b = 0.55$ eV) at points $H$, $H'$, and $D$. We note that, if both Ni$_{\rm Li}$ and Ni$_{\rm Mn}$ occur in the material, the charge-compensation mechanism will be different (see more details below). Mo$_{\rm Mn}^+$ is found to occur in the form of Mo$_{\rm Mn}^0$ (i.e., Mo$^{4+}$ at the Mn site); however, Mo$_{\rm Mn}^0$ is stable only in isolation as this defect configuration becomes unstable toward forming a complex of Mo$_{\rm Mn}^+$ and $\eta^-$ when there is more than one Mo impurity in the supercell (As will be discussed in Sec.~\ref{sec;heavy}, in heavily doped \ce{Li2MnO3}, Mo is also stable as Mo$^{5+}$ and the corresponding defect Mo$_{\rm Mn}^+$ is charge-compensated by $\eta^-$). Under the conditions at point $D$, Mo$_{\rm Mn}^0$ and Mo$_{\rm Mn}^+$$-$$V_{\rm Li}^-$ ($E_b = 0.48$ eV) are degenerate in energy; see Table \ref{tab;models}. Experimentally, Mo was reported to be stable as Mo$^{5+}$.\cite{Ma2014CEJ} Finally, Ru is stable as Ru$_{\rm Mn}^0$ as mentioned earlier, consistent with experiments.\cite{Mori2011JPS,Sathiya2013}

We also carry out explicit calculations for Ni-doped \ce{Li2MnO3} with Ni at both the Li site and the Mn site, using two different models. The first model for the doped system involves one Ni atom at the Li ($2b$) site and one at the Mn ($4g$) site. In this case, the impurities can be described as consisting of Ni$_{\rm Li}^+$ and Ni$_{\rm Mn}^-$. The Ni ion at the Mn site thus adopts the charge state $+3$ and becomes the charge-compensating defect for the Ni$^{2+}$ ion at the Li site. In the second model, two Ni atoms are incorporated at the $2b$ site and one at the $4g$ site. In this case, the impurities are found to occur as a complex of two Ni$_{\rm Li}^+$ and one Ni$_{\rm Mn}^{2-}$; i.e., all the Ni ions are in the $+2$ charge state. The Ni ion at the Mn site thus lowers its charge state further in the presence of the additional Ni$_{\rm Li}^+$ defect. The observations here indicate that charge (and spin) states of a transition-metal impurity can be affected by defect-defect interaction, as also seen in the case of Fe$_{\rm Mn}$ and Mo$_{\rm Mn}$. These results are consistent with those reported in Fig.~\ref{fig;fe}(d), in which all three charge states (+2, +3, and +4) of Ni are found to be stable. We note that the second model can be employed to describe the series of Ni-doped \ce{Li2MnO3} cathode materials Li[Ni$_z$Li$_{1/3-2z/3}$Mn$_{2/3-z/3}$]O$_2$ ($0 \leq z \leq 1/2$), synthesized first by Lu and Dahn\cite{Lu2001} and widely studied and reported in the literature. In our calculations, $z=1/9$, $1/4$, and $1/2$, corresponding to three Ni atoms in the 108-, 48-, and 24-atom cell models, respectively.

\subsection{Electronic structure vis-\`{a}-vis delithiation mechanism in heavy doping}\label{sec;heavy}

Given the lattice site preference and the stable charge and spin states of the impurities reported in the previous section, we can now develop suitable models to describe heavily doped \ce{Li2MnO3} systems and investigate their electronic structure as well as delithiation mechanism. Here, we focus on the Ni, Mo, and Ru impurities, as they are often selected for ion substitution, and make use of the smaller, 24-atom cell size. Since Ni can be incorporated into \ce{Li2MnO3} at the Li site and/or the Mn site, we consider cell models with the following compositions: (i) Li$_{1.75}$Ni$_{0.25}$MnO$_3$ with Ni at the $2b$, $2c$, or $4h$ site, (ii) Li$_2$Mn$_{0.75}$Ni$_{0.25}$O$_3$ with Ni at the $4g$ site, and (iii) Li$_{1.5}$Ni$_{0.75}$Mn$_{0.75}$O$_3$ with two Ni at the $2b$ site and one at the $4g$ site per cell. Model (iii) can also be expressed as Li[Ni$_z$Li$_{1/3-2z/3}$Mn$_{2/3-z/3}$]O$_2$ with $z=1/2$ or LiNi$_{0.5}$Mn$_{0.5}$O$_2$ as often mentioned in the literature. For Ru and Mo, which are found to be incorporated at the $4g$ site, we consider the following compositions: (iv) Li$_2$Mn$_{0.75}$Mo$_{0.25}$O$_3$ and (v) Li$_2$Mn$_{0.75}$Ru$_{0.25}$O$_3$.

\begin{figure}
\centering
\includegraphics[width=8.6cm]{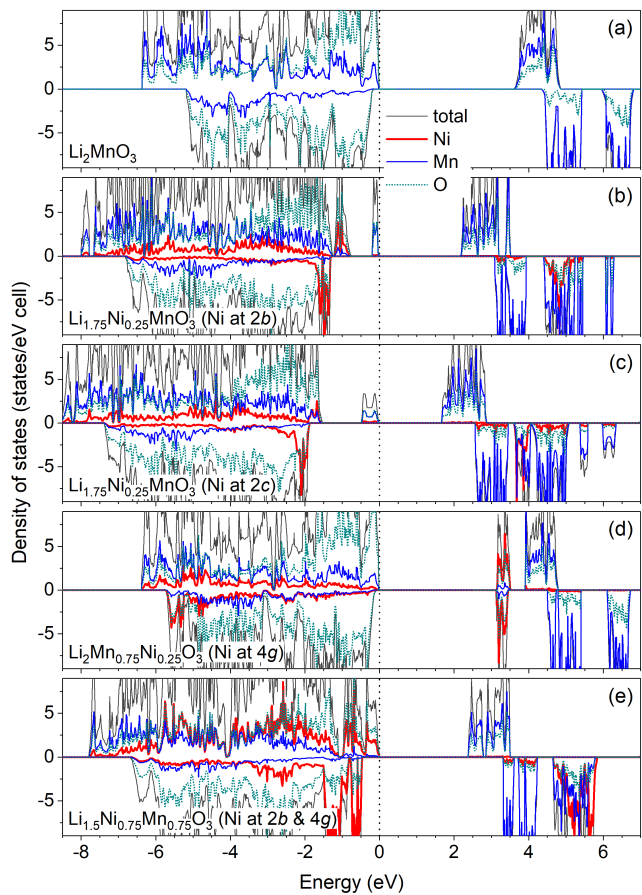}
\vspace{-0.2cm}
\caption{Total and atomic-projected electronic densities of states of (a) undoped \ce{Li2MnO3} and \ce{Li2MnO3} doped with Ni at the (b) $2b$, (c) $2c$, (d) $4g$, or (e) both $2b$ and $4g$ lattice sites. The zero of energy is set to the highest occupied state.}
\label{fig;dos1}
\end{figure}

\begin{figure}
\centering
\includegraphics[width=8.6cm]{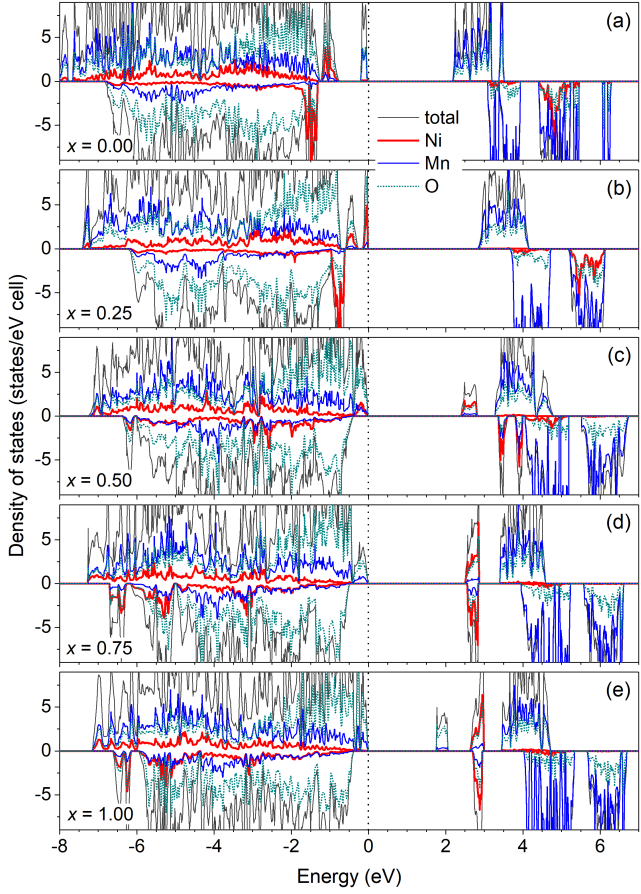}
\vspace{-0.2cm}
\caption{Total and atomic-projected electronic densities of states of Li$_{1.75-x}$Ni$_{0.25}$MnO$_3$ (Ni at the $2b$ site): (a) $x=0.00$, (b) $x=0.25$, (c) $x=0.50$, (d) $x=0.75$, and (e) $x=1.00$. The zero of energy is set to the highest occupied state.}
\label{fig;dos2}
\end{figure}

Figure \ref{fig;dos1} shows the total and atomic-projected electronic density of states (DOS) of undoped and Ni-doped \ce{Li2MnO3}. The electronic structure of the Ni-doped system with Ni at the $4h$ site is very similar to that at the $2c$ site and thus not included in the figure. We find that Ni is stable as Ni$^{2+}$ at the Li site (either $2b$, $2c$, or $4h$) and charge-compensated by Mn$^{3+}$ formed at the Mn site (i.e., $\eta^-$), in agreement with the results obtained in the dilute doping limit. On the other hand, Ni is stable as Ni$^{4+}$ at the $4g$ site and all Ni in Li$_{1.5}$Ni$_{0.75}$Mn$_{0.75}$O$_3$ are stable as Ni$^{2+}$ at the $2b$ and $4g$ sites as reported earlier; $\eta^-$ is not found in these two cases. The electronic structure, particularly near the band edges, is strongly dependent on the lattice site of the dopant. We are mostly interested in the nature of the electronic structure near the VBM as it provides information on the mechanism for lithium extraction.\cite{Hoang2011CM,Hoang2015PRA} The results show that the VBM of Li$_{1.75}$Ni$_{0.25}$MnO$_3$ is composed predominantly of the $3d$ states from the Mn$^{3+}$ ion; see Fig.~\ref{fig;dos1}(b) and \ref{fig;dos1}(c). In Li$_2$Mn$_{0.75}$Ni$_{0.25}$O$_3$, the VBM is predominantly the O $2p$ states, see Fig.~\ref{fig;dos1}(d), similar to that in undoped \ce{Li2MnO3} [Fig.~\ref{fig;dos1}(a)]. This is because when doped at the $4g$ site only, Ni is stable as Ni$^{4+}$ and the Ni $3d$ states are in the conduction band. Finally, in Li$_{1.5}$Ni$_{0.75}$Mn$_{0.75}$O$_3$, the VBM is predominantly the $3d$ states from the Ni$^{2+}$ ions; see Fig.~\ref{fig;dos1}(e). Note that, though the atomic-decomposed DOS in the figures shows the total contribution from the materials' constituting elements, our analysis is based on contributions from individual atoms which contain more details on the nature of the electronic structure. 

The features observed in the electronic structure of the Ni-doped \ce{Li2MnO3} systems will result in different delithiation mechanisms. As Li in the material already donates its electron to the lattice and becomes Li$^+$, lithium extraction involves removing the Li$^{+}$ ion and one electron from the lattice (specifically, from the highest occupied state).\cite{Hoang2011CM,Hoang2015PRA} The electronic structures in Fig.~\ref{fig;dos1} indicate that, upon delithiation, Mn$^{3+}$ in Li$_{1.75}$Ni$_{0.25}$MnO$_3$ will be the first to be oxidized; Li$_2$Mn$_{0.75}$Ni$_{0.25}$O$_3$ will have the same delithiation mechanism as undoped \ce{Li2MnO3} which was found to involve the oxidation of O$^{2-}$ to O$^-$;\cite{Hoang2015PRA} and the oxidation will occur first with Ni$^{2+}$ in Li$_{1.5}$Ni$_{0.75}$Mn$_{0.75}$O$_3$. The mechanism for subsequent stages of delithiation will be determined by the electronic structure of the partially delithiated compounds. 

\begin{figure}
\centering
\includegraphics[width=8.6cm]{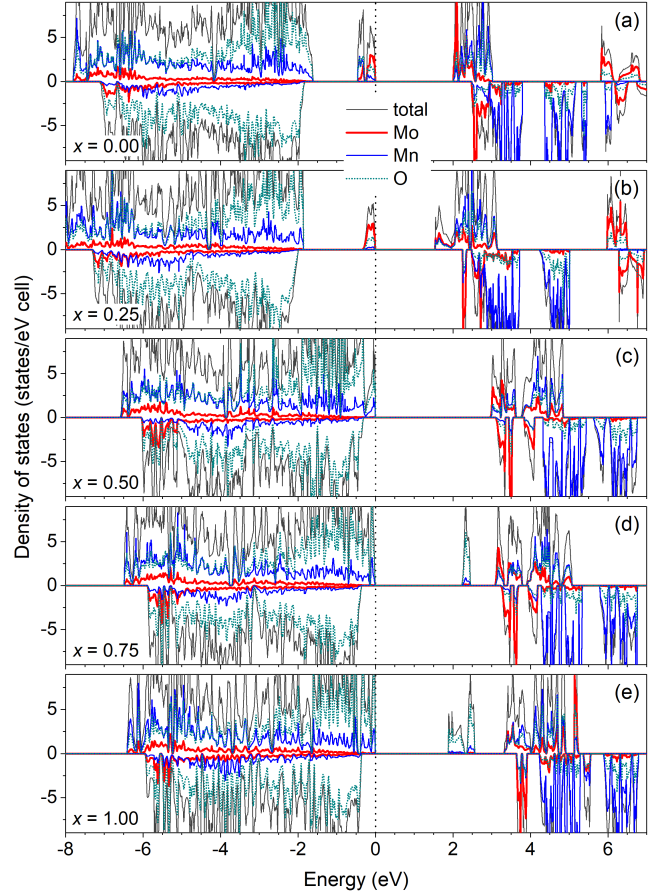}
\vspace{-0.2cm}
\caption{Total and atomic-projected electronic densities of states of Li$_{2-x}$Mn$_{0.75}$Mo$_{0.25}$O$_3$ (Mo at the $4g$ site): (a) $x=0.00$, (b) $x=0.25$, (c) $x=0.50$, (d) $x=0.75$, and (e) $x=1.00$. The zero of energy is set to the highest occupied state.}
\label{fig;dos3}
\end{figure}

\begin{figure}
\centering
\includegraphics[width=8.6cm]{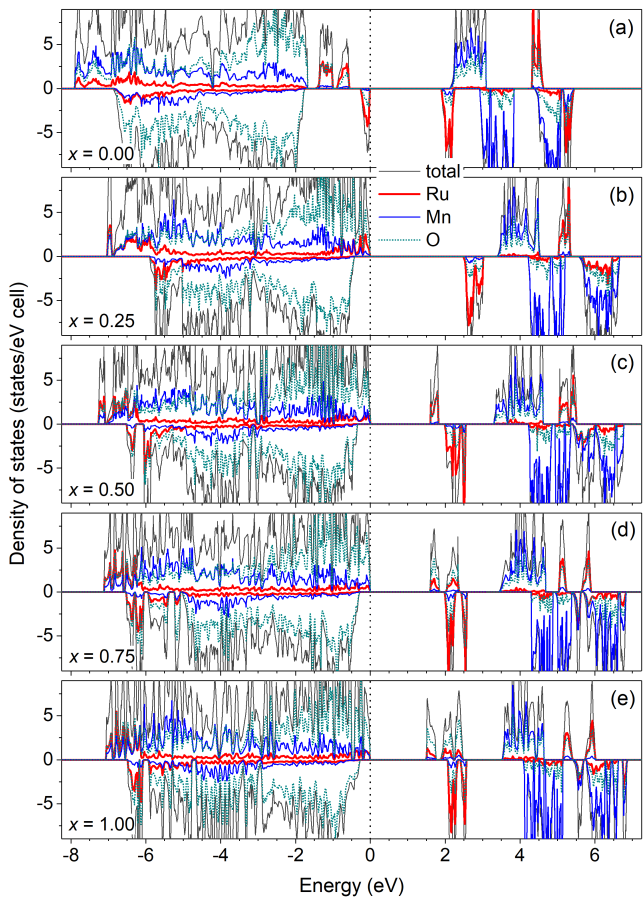}
\vspace{-0.2cm}
\caption{Total and atomic-projected electronic densities of states of Li$_{2-x}$Mn$_{0.75}$Ru$_{0.25}$O$_3$ (Ru at the $4g$ site): (a) $x=0.00$, (b) $x=0.25$, (c) $x=0.50$, (d) $x=0.75$, and (e) $x=1.00$. The zero of energy is set to the highest occupied state.}
\label{fig;dos4}
\end{figure} 

In order to demonstrate how the electronic structure of a Ni-doped \ce{Li2MnO3} material changes upon delithiation, we show in Fig.~\ref{fig;dos2} the DOS of Li$_{1.75-x}$Ni$_{0.25}$MnO$_3$ (with the dopant at the $2b$ site only) for different degrees of delithiation. At $x=0$, the valence-band top is predominantly the $3d$ states from the Mn$^{3+}$ ion; see Fig.~\ref{fig;dos2}(a). Once an electron is removed from Mn$^{3+}$ and the ion is oxidized to Mn$^{4+}$, at $x=0.25$, the Mn $3d$ states are empty and pushed to the conduction band; the valence-band top is now predominantly the $3d$ states from the Ni$^{2+}$ ion; see Fig.~\ref{fig;dos2}(b). Next, an electron is removed from the Ni$^{2+}$ $3d$ states and the ion is oxidized to Ni$^{3+}$; see Fig.~\ref{fig;dos2}(c). The Ni$^{3+}$ is then oxidized to Ni$^{4+}$ at $x=0.75$; see Fig.~\ref{fig;dos2}(d). Now, Ni and Mn are all in the charge state $+4$ and the valence-band top is predominantly the O $2p$ states which is similar to that of undoped \ce{Li2MnO3}. Lithium extraction beyond this point involves the oxidation of O$^{2-}$ to O$^{-}$. For example, at $x=1.00$, Li$_{1.75-x}$Ni$_{0.25}$MnO$_3$ has electronic states at $\sim$2 eV that are associated with the oxygen hole polaron $\eta_{\rm O}^+$ (i.e., O$^-$); see Fig.~\ref{fig;dos2}(e). 

We note that, since the nature of the electronic structure near the band edges is the same for \ce{Li2MnO3} doped with Ni at the $2b$ site, the $2c$ site, or the $4h$ site, the delithiation mechanisms are the same in these systems.    

Figure \ref{fig;dos3} shows the DOS of Li$_{2-x}$Mn$_{0.75}$Mo$_{0.25}$O$_3$ (with the dopant incorporated at the $4g$ site) for different degrees of delithiation. Mo is stable as Mo$^{5+}$ and charge-compensated by Mn$^{3+}$ as mentioned in Sec.~\ref{sec;dilute}. At $x=0$, the valence-band top is predominantly the Mn$^{3+}$ $3d$ and Mo$^{5+}$ $4d$ states. We find that an electron is removed first from Mn$^{3+}$ and the ion is oxidized to Mn$^{4+}$; the valence-band top becomes predominantly the Mo$^{5+}$ $4d$ states after the removal; see Fig.~\ref{fig;dos3}(b). Subsequent removal of lithium turns Mo$^{5+}$ into Mo$^{6+}$ and the Mo $4d$ states become empty and are pushed up in energy; see Fig.~\ref{fig;dos3}(c). From this point onward, lithium extraction involves the oxidation of O$^{2-}$ to O$^-$. For example, the DOS peak at about 2.5 eV in Fig.~\ref{fig;dos3}(d) is associated with an oxygen hole polaron $\eta_{\rm O}^+$ resulted from the O$^{2-/-}$ redox reaction. At $x=1.00$, the DOS shows two peaks at $\sim$2.0$-$2.5 eV, corresponding to $\eta_{\rm O}^+$ defects present in the partially delithiated compound; see Fig.~\ref{fig;dos3}(e).         

\begin{figure}
\centering
\includegraphics[width=8.6cm]{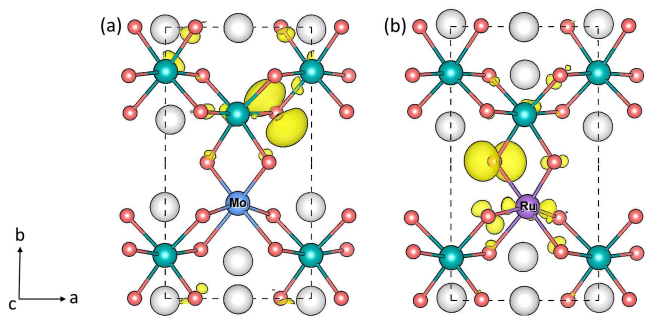}
\vspace{-0.2cm}
\caption{Charge densities associated with the oxygen hole polaron $\eta_{\rm O}^+$ (i.e., O$^-$) in Li$_{2-x}$Mn$_{0.75}$M$_{0.25}$O$_3$ ($x=0.75$): (a) M = Mo and (b) M = Ru. The iso-value for the isosurface (yellow) is set to 0.05 e/{\AA}$^3$. Large gray spheres are Li, medium blue spheres are Mn, and small red spheres are O.}
\label{fig;eta}
\end{figure}

Figure \ref{fig;dos4} shows the DOS of Li$_{2-x}$Mn$_{0.75}$Ru$_{0.25}$O$_3$ (with the dopant at the $4g$ site) for different degrees of delithiation. Similar to the other doped \ce{Li2MnO3} systems discussed earlier, the electronic structure of the Ru-doped one, particularly the nature of the electronic states near the band edges, changes as the delithiation process progresses. At $x=0$, the top of the valence band is predominantly the Ru $4d$ states; see Fig.~\ref{fig;dos4}(a). The first two stages of delithition involve oxidation of Ru$^{4+}$ to Ru$^{5+}$, see Fig.~\ref{fig;dos4}(b), and then from Ru$^{5+}$ to Ru$^{6+}$, see Fig.~\ref{fig;dos4}(c). For $x\geq 0.75$, oxygen in the material becomes oxidized, leading to the formation of oxygen hole polarons $\eta_{\rm O}^+$. The DOS peaks (at the bottom of the conduction band) associated with $\eta_{\rm O}^+$ can be observed in Figs.~\ref{fig;dos4}(d) and \ref{fig;dos4}(e). There is a strong mixing of the $\eta_{\rm O}^+$ states and Ru $4d$ states, a feature that is not seen in the Ni- or Mo-doped system, indicating strong Ru$-$O interactions. Figure \ref{fig;eta} shows charge densities associated with the oxygen hole polaron $\eta_{\rm O}^+$ in Li$_{2-x}$Mn$_{0.75}$M$_{0.25}$O$_3$ ($x=0.75$, M = Mo, Ru); $\eta_{\rm O}^+$ is found to be located between two Mn in one compound and between Mn and Ru in the other.  

We note that the doped \ce{Li2MnO3} materials all exhibit non-metallic behavior and they remain non-metallic upon delithiation. This observation is consistent with our analysis of defect physics in \ce{Li2MnO3} reported previously\cite{Hoang2015PRA} according to which the material cannot be doped $n$- or $p$-type like a conventional semiconductor; any attempt to deliberately shift the Fermi level to the VBM or CBM will result in having intrinsic point defects formed spontaneously to counteract the effect of shifting. Our results are thus in contrast to those reported by other research groups, often obtained in DFT$+U$ calculations, according to which partially delithiated (and even some fully lithiated) \ce{Li2MnO3}-based materials show metallic behavior.\cite{Yang2017AFM,Gao2014JMCA,Gao2015CM} Like in the case of undoped \ce{Li2MnO3} discussed earlier (Sec.~\ref{sec;method}), experiments should be performed to confirm the nature of bulk electronic conduction.

\subsection{Doping as enabler of the anionic oxidation reaction during delithiation}

\begin{figure}
\centering
\includegraphics[width=8.6cm]{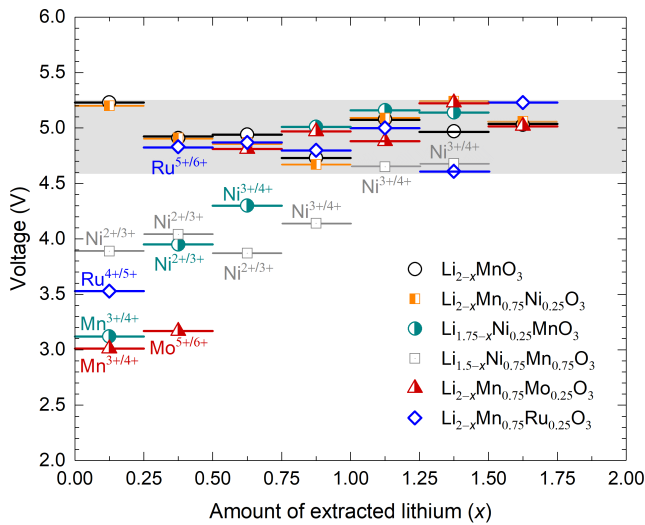}
\vspace{-0.2cm}
\caption{Voltage profiles of undoped \ce{Li2MnO3} (black circles) and \ce{Li2MnO3} doped with Ni at the $4g$ site (orange half-filled squares), the $2b$ site (green half-filled circles), or both the $2b$ and $4g$ sites (gray squares); Mo (red half-filled triangles) or Ru (blue diamonds) at the $4g$ site. Each voltage point is the average voltage between the two Li concentrations used to calculate it. Unmarked voltage points in the shaded area are those associated with the redox activity on oxygen.}
\label{fig;voltage}
\end{figure}

Figure \ref{fig;voltage} shows the voltage profiles of the undoped and doped \ce{Li2MnO3} materials. The redox couples associated with different voltage points are indicated in the figure; unmarked points in the shaded region are associated with the redox activity on oxygen, which can be assigned to the O$^{2-/-}$ couple, at least nominally. All these redox couples are determined by observing changes on the magnetic moment and charge density induced by lithium removal. We find that the voltage of undoped \ce{Li2MnO3} is high at the beginning but lower after the first stage of delithiation. The voltage profile of Li$_{2-x}$Ni$_{0.75}$Mn$_{0.75}$O$_3$ (with Ni at the $4g$ site only) follow closely that of undoped \ce{Li2MnO3}, which is expected, given that the nature of the valence band of the two systems is very similar. In Li$_{1.5-x}$Ni$_{0.75}$Mn$_{0.75}$O$_3$, only the Ni ions are oxidized because they provide already enough electrons for delithiation. We note that, if the Ni concentration is lower, i.e., Li[Ni$_z$Li$_{1/3-2z/3}$Mn$_{2/3-z/3}$]O$_2$ with $z<1/2$, oxygen would be oxidized at high degrees of delithiation, as also observed in experiments.\cite{Luo2016JACS} This is, indeed, the case, as seen in our calculations using the 48-atom supercells which correspond the composition with $z=1/4$. For the other doped \ce{Li2MnO3} systems, oxidation occurs first with the transition-metal ions whereas contributions from the oxygen usually occurs at higher voltages. These results are thus consistent with our analysis of the calculated electronic structure reported earlier. 

As discussed in Ref.~\citenum{Hoang2015PRA}, the difficulty in activating the undoped \ce{Li2MnO3} material electrochemically is likely due to the high extraction voltage and a lack of percolation pathways for efficient electronic conduction in the bulk at the onset of delithiation. Other delithiation mechanisms, including non-intrinsic ones, may thus become essential in the early stages of the delithiation process; these include simultaneous removal of lithium and oxygen at the surfaces and/or interfaces (which leads to oxygen release as often discussed in the literature), oxidation of preexisting Mn$^{3+}$ defects in the material, and/or oxidation of the electrolyte and exchange of H$^+$ for Li$^+$.\cite{Hoang2015PRA} For example, when a Li is replaced by H, the resulted impurity (nominally denoted as H$_{\rm Li}^0$) can be regarded as a defect complex of a positively charged hydrogen interstitial H$_i^+$ and a negatively charged lithium vacancy $V_{\rm Li}^-$. The exchange of H$^+$ for Li$^+$ thus also results in the formation of $V_{\rm Li}^-$ that is needed to stabilize $\eta_{\rm O}^+$ and to establish the percolation pathways. Indeed, we find that $\eta_{\rm O}^+$ forms upon electron removal from H-doped (partially protonated) \ce{Li2MnO3}. The average voltage between Li$_{1.75}$H$_{0.25}$MnO$_3$ and Li$_{1.50}$H$_{0.25}$MnO$_3$ is calculated to be 5.06 V, which is lower than that obtained for undoped \ce{Li2MnO3} during the first stage of delithiation.

In Ni-, Mo-, or Ru-doped \ce{Li2MnO3}, charge compensation and bulk electronic conduction in the initial stages of delithiation are provided by the electrochemically active transition-metal ions. The lower voltages associated with these non-oxygen redox couples also make it easier for lithium removal. The mechanism involving the O$^{2-/-}$ redox couple is expected to be more efficient in the later stages where the $V_{\rm Li}^-$ concentration is high and hence bulk electronic transport via $\eta_{\rm O}^+$ becomes possible.\cite{Hoang2015PRA}       

Finally, we note that the average voltage associated with the oxide redox reaction in the intermediate stages of delithiation is $\sim$4.9 V; see Fig.~\ref{fig;voltage}. This value is thus only about 0.4 V higher than the $\sim$4.5 V plateau\cite{Lu2002JES,Koga2014JPCC,Luo2016JACS} often observed in \ce{Li2MnO3}-based materials during first charge and attributed to the oxidation of oxygen in the bulk lattice. The voltage difference is well within that often seen in voltage calculations; see, e.g., Ref.~\citenum{Chevrier2010}. 

\section{Conclusions}

A detailed study of doping in Li-rich cathode material \ce{Li2MnO3} has been carried out using hybrid density-functional calculations. We consider the impurities in both the dilute doping limit and heavy doping with attention to charge-compensation mechanisms and the effects of defect-defect interaction on the stable charge and spin states of the transition-metal ions. Al, Fe, Mo, and Ru impurities are found to be energetically most favorable when incorporated at the Mn lattice site, whereas Mg is most favorable when doped at the Li sites; Ni can be incorporated at the Li site and/or the Mn site. There is a strong interplay between the lattice site preference and charge and spin states of the dopant, the electronic structure of the doped material, and the mechanism for lithium extraction. The calculated electronic structures and voltage profiles provide specific information on the delithiation mechanisms, including the order in which the redox couples are activated. In these mechanisms, electrochemically active transition-metal ions provide charge-compensation and bulk electronic conduction mechanisms in the initial stages of the delithiation process and hence enable the mechanism involving the oxidation of the lattice oxygen in the later stages. 

\begin{acknowledgments}

The calculations were carried out using computing resources at the Center for Computationally Assisted Science and Technology at North Dakota State University.

\end{acknowledgments}


\begin{thebibliography}{46}%
\makeatletter
\providecommand \@ifxundefined [1]{%
 \@ifx{#1\undefined}
}%
\providecommand \@ifnum [1]{%
 \ifnum #1\expandafter \@firstoftwo
 \else \expandafter \@secondoftwo
 \fi
}%
\providecommand \@ifx [1]{%
 \ifx #1\expandafter \@firstoftwo
 \else \expandafter \@secondoftwo
 \fi
}%
\providecommand \natexlab [1]{#1}%
\providecommand \enquote  [1]{``#1''}%
\providecommand \bibnamefont  [1]{#1}%
\providecommand \bibfnamefont [1]{#1}%
\providecommand \citenamefont [1]{#1}%
\providecommand \href@noop [0]{\@secondoftwo}%
\providecommand \href [0]{\begingroup \@sanitize@url \@href}%
\providecommand \@href[1]{\@@startlink{#1}\@@href}%
\providecommand \@@href[1]{\endgroup#1\@@endlink}%
\providecommand \@sanitize@url [0]{\catcode `\\12\catcode `\$12\catcode
  `\&12\catcode `\#12\catcode `\^12\catcode `\_12\catcode `\%12\relax}%
\providecommand \@@startlink[1]{}%
\providecommand \@@endlink[0]{}%
\providecommand \url  [0]{\begingroup\@sanitize@url \@url }%
\providecommand \@url [1]{\endgroup\@href {#1}{\urlprefix }}%
\providecommand \urlprefix  [0]{URL }%
\providecommand \Eprint [0]{\href }%
\providecommand \doibase [0]{http://dx.doi.org/}%
\providecommand \selectlanguage [0]{\@gobble}%
\providecommand \bibinfo  [0]{\@secondoftwo}%
\providecommand \bibfield  [0]{\@secondoftwo}%
\providecommand \translation [1]{[#1]}%
\providecommand \BibitemOpen [0]{}%
\providecommand \bibitemStop [0]{}%
\providecommand \bibitemNoStop [0]{.\EOS\space}%
\providecommand \EOS [0]{\spacefactor3000\relax}%
\providecommand \BibitemShut  [1]{\csname bibitem#1\endcsname}%
\let\auto@bib@innerbib\@empty
\bibitem [{\citenamefont {Thackeray}\ \emph {et~al.}(2005)\citenamefont
  {Thackeray}, \citenamefont {Johnson}, \citenamefont {Vaughey}, \citenamefont
  {Li},\ and\ \citenamefont {Hackney}}]{Thackeray2005JMC}%
  \BibitemOpen
  \bibfield  {author} {\bibinfo {author} {\bibfnamefont {M.~M.}\ \bibnamefont
  {Thackeray}}, \bibinfo {author} {\bibfnamefont {C.~S.}\ \bibnamefont
  {Johnson}}, \bibinfo {author} {\bibfnamefont {J.~T.}\ \bibnamefont
  {Vaughey}}, \bibinfo {author} {\bibfnamefont {N.}~\bibnamefont {Li}}, \ and\
  \bibinfo {author} {\bibfnamefont {S.~A.}\ \bibnamefont {Hackney}},\
  }\bibfield  {title} {\enquote {\bibinfo {title} {Advances in manganese-oxide
  {'}composite{'} electrodes for lithium-ion batteries},}\ }\href {\doibase
  10.1039/B417616M} {\bibfield  {journal} {\bibinfo  {journal} {J. Mater.
  Chem.}\ }\textbf {\bibinfo {volume} {15}},\ \bibinfo {pages} {2257--2267}
  (\bibinfo {year} {2005})}\BibitemShut {NoStop}%
\bibitem [{\citenamefont {Croguennec}\ and\ \citenamefont
  {Palacin}(2015)}]{Croguennec2015JACS}%
  \BibitemOpen
  \bibfield  {author} {\bibinfo {author} {\bibfnamefont {L.}~\bibnamefont
  {Croguennec}}\ and\ \bibinfo {author} {\bibfnamefont {M.~R.}\ \bibnamefont
  {Palacin}},\ }\bibfield  {title} {\enquote {\bibinfo {title} {{Recent
  Achievements on Inorganic Electrode Materials for Lithium-Ion Batteries}},}\
  }\href {\doibase 10.1021/ja507828x} {\bibfield  {journal} {\bibinfo
  {journal} {J. Am. Chem. Soc.}\ }\textbf {\bibinfo {volume} {137}},\ \bibinfo
  {pages} {3140--3156} (\bibinfo {year} {2015})}\BibitemShut {NoStop}%
\bibitem [{\citenamefont {Massarotti}\ \emph {et~al.}(1997)\citenamefont
  {Massarotti}, \citenamefont {Bini}, \citenamefont {Capsoni}, \citenamefont
  {Altomare},\ and\ \citenamefont {Moliterni}}]{Massarotti1997}%
  \BibitemOpen
  \bibfield  {author} {\bibinfo {author} {\bibfnamefont {V.}~\bibnamefont
  {Massarotti}}, \bibinfo {author} {\bibfnamefont {M.}~\bibnamefont {Bini}},
  \bibinfo {author} {\bibfnamefont {D.}~\bibnamefont {Capsoni}}, \bibinfo
  {author} {\bibfnamefont {A.}~\bibnamefont {Altomare}}, \ and\ \bibinfo
  {author} {\bibfnamefont {A.~G.~G.}\ \bibnamefont {Moliterni}},\ }\bibfield
  {title} {\enquote {\bibinfo {title} {{{\it Ab Initio} Structure Determination
  of Li${\sb 2}$MnO${\sb 3}$ from X-ray Powder Diffraction Data}},}\ }\href
  {\doibase 10.1107/S0021889896012460} {\bibfield  {journal} {\bibinfo
  {journal} {J. Appl. Cryst.}\ }\textbf {\bibinfo {volume} {30}},\ \bibinfo
  {pages} {123--127} (\bibinfo {year} {1997})}\BibitemShut {NoStop}%
\bibitem [{\citenamefont {Kalyani}\ \emph {et~al.}(1999)\citenamefont
  {Kalyani}, \citenamefont {Chitra}, \citenamefont {Mohan},\ and\ \citenamefont
  {Gopukumar}}]{Kalyani1999}%
  \BibitemOpen
  \bibfield  {author} {\bibinfo {author} {\bibfnamefont {P.}~\bibnamefont
  {Kalyani}}, \bibinfo {author} {\bibfnamefont {S.}~\bibnamefont {Chitra}},
  \bibinfo {author} {\bibfnamefont {T.}~\bibnamefont {Mohan}}, \ and\ \bibinfo
  {author} {\bibfnamefont {S.}~\bibnamefont {Gopukumar}},\ }\bibfield  {title}
  {\enquote {\bibinfo {title} {Lithium metal rechargeable cells using
  \ce{Li2MnO3} as the positive electrode},}\ }\href {\doibase
  http://dx.doi.org/10.1016/S0378-7753(99)00066-X} {\bibfield  {journal}
  {\bibinfo  {journal} {J. Power Sources}\ }\textbf {\bibinfo {volume} {80}},\
  \bibinfo {pages} {103--106} (\bibinfo {year} {1999})}\BibitemShut {NoStop}%
\bibitem [{\citenamefont {Sathiya}\ \emph {et~al.}(2013)\citenamefont
  {Sathiya}, \citenamefont {Ramesha}, \citenamefont {Rousse}, \citenamefont
  {Foix}, \citenamefont {Gonbeau}, \citenamefont {Prakash}, \citenamefont
  {Doublet}, \citenamefont {Hemalatha},\ and\ \citenamefont
  {Tarascon}}]{Sathiya2013}%
  \BibitemOpen
  \bibfield  {author} {\bibinfo {author} {\bibfnamefont {M.}~\bibnamefont
  {Sathiya}}, \bibinfo {author} {\bibfnamefont {K.}~\bibnamefont {Ramesha}},
  \bibinfo {author} {\bibfnamefont {G.}~\bibnamefont {Rousse}}, \bibinfo
  {author} {\bibfnamefont {D.}~\bibnamefont {Foix}}, \bibinfo {author}
  {\bibfnamefont {D.}~\bibnamefont {Gonbeau}}, \bibinfo {author} {\bibfnamefont
  {A.~S.}\ \bibnamefont {Prakash}}, \bibinfo {author} {\bibfnamefont {M.~L.}\
  \bibnamefont {Doublet}}, \bibinfo {author} {\bibfnamefont {K.}~\bibnamefont
  {Hemalatha}}, \ and\ \bibinfo {author} {\bibfnamefont {J.-M.}\ \bibnamefont
  {Tarascon}},\ }\bibfield  {title} {\enquote {\bibinfo {title} {{High
  Performance Li$_2$Ru$_{1-y}$Mn$_y$O$_3$ ($0.2 \leq y \leq 0.8$) Cathode
  Materials for Rechargeable Lithium-Ion Batteries: Their Understanding}},}\
  }\href {\doibase 10.1021/cm400193m} {\bibfield  {journal} {\bibinfo
  {journal} {Chem. Mater.}\ }\textbf {\bibinfo {volume} {25}},\ \bibinfo
  {pages} {1121--1131} (\bibinfo {year} {2013})}\BibitemShut {NoStop}%
\bibitem [{\citenamefont {Hoang}(2015)}]{Hoang2015PRA}%
  \BibitemOpen
  \bibfield  {author} {\bibinfo {author} {\bibfnamefont {K.}~\bibnamefont
  {Hoang}},\ }\bibfield  {title} {\enquote {\bibinfo {title} {{Defect Physics,
  Delithiation Mechanism, and Electronic and Ionic Conduction in Layered
  Lithium Manganese Oxide Cathode Materials}},}\ }\href {\doibase
  10.1103/PhysRevApplied.3.024013} {\bibfield  {journal} {\bibinfo  {journal}
  {Phys. Rev. Applied}\ }\textbf {\bibinfo {volume} {3}},\ \bibinfo {pages}
  {024013} (\bibinfo {year} {2015})}\BibitemShut {NoStop}%
\bibitem [{\citenamefont {Luo}\ \emph {et~al.}(2016{\natexlab{a}})\citenamefont
  {Luo}, \citenamefont {Roberts}, \citenamefont {Hao}, \citenamefont
  {Guerrini}, \citenamefont {Pickup}, \citenamefont {Liu}, \citenamefont
  {Edstrom}, \citenamefont {Guo}, \citenamefont {Chadwick}, \citenamefont
  {Duda},\ and\ \citenamefont {Bruce}}]{Luo2016NC}%
  \BibitemOpen
  \bibfield  {author} {\bibinfo {author} {\bibfnamefont {K.}~\bibnamefont
  {Luo}}, \bibinfo {author} {\bibfnamefont {M.~R.}\ \bibnamefont {Roberts}},
  \bibinfo {author} {\bibfnamefont {R.}~\bibnamefont {Hao}}, \bibinfo {author}
  {\bibfnamefont {N.}~\bibnamefont {Guerrini}}, \bibinfo {author}
  {\bibfnamefont {D.~M.}\ \bibnamefont {Pickup}}, \bibinfo {author}
  {\bibfnamefont {Y.-S.}\ \bibnamefont {Liu}}, \bibinfo {author} {\bibfnamefont
  {K.}~\bibnamefont {Edstrom}}, \bibinfo {author} {\bibfnamefont
  {J.}~\bibnamefont {Guo}}, \bibinfo {author} {\bibfnamefont {A.~V.}\
  \bibnamefont {Chadwick}}, \bibinfo {author} {\bibfnamefont {L.~C.}\
  \bibnamefont {Duda}}, \ and\ \bibinfo {author} {\bibfnamefont {P.~G.}\
  \bibnamefont {Bruce}},\ }\bibfield  {title} {\enquote {\bibinfo {title}
  {{Charge-compensation in 3d-transition-metal-oxide intercalation cathodes
  through the generation of localized electron holes on oxygen}},}\ }\href
  {\doibase 10.1038/nchem.2471} {\bibfield  {journal} {\bibinfo  {journal}
  {Nature Chem.}\ }\textbf {\bibinfo {volume} {8}},\ \bibinfo {pages}
  {684--691} (\bibinfo {year} {2016}{\natexlab{a}})}\BibitemShut {NoStop}%
\bibitem [{\citenamefont {Yu}\ \emph {et~al.}(2016)\citenamefont {Yu},
  \citenamefont {Wang}, \citenamefont {Fu}, \citenamefont {Wang}, \citenamefont
  {Cai}, \citenamefont {Liu}, \citenamefont {Lu}, \citenamefont {Wang},
  \citenamefont {Wang}, \citenamefont {Ren},\ and\ \citenamefont
  {Yang}}]{Yu2016JMCA}%
  \BibitemOpen
  \bibfield  {author} {\bibinfo {author} {\bibfnamefont {R.}~\bibnamefont
  {Yu}}, \bibinfo {author} {\bibfnamefont {X.}~\bibnamefont {Wang}}, \bibinfo
  {author} {\bibfnamefont {Y.}~\bibnamefont {Fu}}, \bibinfo {author}
  {\bibfnamefont {L.}~\bibnamefont {Wang}}, \bibinfo {author} {\bibfnamefont
  {S.}~\bibnamefont {Cai}}, \bibinfo {author} {\bibfnamefont {M.}~\bibnamefont
  {Liu}}, \bibinfo {author} {\bibfnamefont {B.}~\bibnamefont {Lu}}, \bibinfo
  {author} {\bibfnamefont {G.}~\bibnamefont {Wang}}, \bibinfo {author}
  {\bibfnamefont {D.}~\bibnamefont {Wang}}, \bibinfo {author} {\bibfnamefont
  {Q.}~\bibnamefont {Ren}}, \ and\ \bibinfo {author} {\bibfnamefont
  {X.}~\bibnamefont {Yang}},\ }\bibfield  {title} {\enquote {\bibinfo {title}
  {Effect of magnesium doping on properties of lithium-rich layered oxide
  cathodes based on a one-step co-precipitation strategy},}\ }\href {\doibase
  10.1039/C6TA00370B} {\bibfield  {journal} {\bibinfo  {journal} {J. Mater.
  Chem. A}\ }\textbf {\bibinfo {volume} {4}},\ \bibinfo {pages} {4941--4951}
  (\bibinfo {year} {2016})}\BibitemShut {NoStop}%
\bibitem [{\citenamefont {Xiang}\ and\ \citenamefont
  {Wu}(2017)}]{Xiang2017Ionics}%
  \BibitemOpen
  \bibfield  {author} {\bibinfo {author} {\bibfnamefont {Y.}~\bibnamefont
  {Xiang}}\ and\ \bibinfo {author} {\bibfnamefont {X.}~\bibnamefont {Wu}},\
  }\bibfield  {title} {\enquote {\bibinfo {title} {{Enhanced electrochemical
  performances of \ce{Li2MnO3} cathode materials by Al doping}},}\ }\href
  {\doibase 10.1007/s11581-017-2189-4} {\bibfield  {journal} {\bibinfo
  {journal} {Ionics}\ } (\bibinfo {year} {2017}),\
  10.1007/s11581-017-2189-4}\BibitemShut {NoStop}%
\bibitem [{\citenamefont {Yuge}\ \emph {et~al.}(2017)\citenamefont {Yuge},
  \citenamefont {Kuroshima}, \citenamefont {Toda}, \citenamefont {Miyazaki},
  \citenamefont {Tabuchi}, \citenamefont {Doumae}, \citenamefont {Shibuya},\
  and\ \citenamefont {Tamura}}]{Yuge2017JPS}%
  \BibitemOpen
  \bibfield  {author} {\bibinfo {author} {\bibfnamefont {R.}~\bibnamefont
  {Yuge}}, \bibinfo {author} {\bibfnamefont {S.}~\bibnamefont {Kuroshima}},
  \bibinfo {author} {\bibfnamefont {A.}~\bibnamefont {Toda}}, \bibinfo {author}
  {\bibfnamefont {T.}~\bibnamefont {Miyazaki}}, \bibinfo {author}
  {\bibfnamefont {M.}~\bibnamefont {Tabuchi}}, \bibinfo {author} {\bibfnamefont
  {K.}~\bibnamefont {Doumae}}, \bibinfo {author} {\bibfnamefont
  {H.}~\bibnamefont {Shibuya}}, \ and\ \bibinfo {author} {\bibfnamefont
  {N.}~\bibnamefont {Tamura}},\ }\bibfield  {title} {\enquote {\bibinfo {title}
  {Structural and electrochemical properties of iron- and nickel-substituted
  \ce{Li2MnO3} cathodes in charged and discharged states},}\ }\href {\doibase
  http://dx.doi.org/10.1016/j.jpowsour.2017.08.049} {\bibfield  {journal}
  {\bibinfo  {journal} {J. Power Sources}\ }\textbf {\bibinfo {volume} {365}},\
  \bibinfo {pages} {117--125} (\bibinfo {year} {2017})}\BibitemShut {NoStop}%
\bibitem [{\citenamefont {Luo}\ \emph {et~al.}(2016{\natexlab{b}})\citenamefont
  {Luo}, \citenamefont {Roberts}, \citenamefont {Guerrini}, \citenamefont
  {Tapia-Ruiz}, \citenamefont {Hao}, \citenamefont {Massel}, \citenamefont
  {Pickup}, \citenamefont {Ramos}, \citenamefont {Liu}, \citenamefont {Guo},
  \citenamefont {Chadwick}, \citenamefont {Duda},\ and\ \citenamefont
  {Bruce}}]{Luo2016JACS}%
  \BibitemOpen
  \bibfield  {author} {\bibinfo {author} {\bibfnamefont {K.}~\bibnamefont
  {Luo}}, \bibinfo {author} {\bibfnamefont {M.~R.}\ \bibnamefont {Roberts}},
  \bibinfo {author} {\bibfnamefont {N.}~\bibnamefont {Guerrini}}, \bibinfo
  {author} {\bibfnamefont {N.}~\bibnamefont {Tapia-Ruiz}}, \bibinfo {author}
  {\bibfnamefont {R.}~\bibnamefont {Hao}}, \bibinfo {author} {\bibfnamefont
  {F.}~\bibnamefont {Massel}}, \bibinfo {author} {\bibfnamefont {D.~M.}\
  \bibnamefont {Pickup}}, \bibinfo {author} {\bibfnamefont {S.}~\bibnamefont
  {Ramos}}, \bibinfo {author} {\bibfnamefont {Y.-S.}\ \bibnamefont {Liu}},
  \bibinfo {author} {\bibfnamefont {J.}~\bibnamefont {Guo}}, \bibinfo {author}
  {\bibfnamefont {A.~V.}\ \bibnamefont {Chadwick}}, \bibinfo {author}
  {\bibfnamefont {L.~C.}\ \bibnamefont {Duda}}, \ and\ \bibinfo {author}
  {\bibfnamefont {P.~G.}\ \bibnamefont {Bruce}},\ }\bibfield  {title} {\enquote
  {\bibinfo {title} {{Anion Redox Chemistry in the Cobalt Free 3d Transition
  Metal Oxide Intercalation Electrode
  Li[Li$_{0.2}$Ni$_{0.2}$Mn$_{0.6}$]O$_2$}},}\ }\href {\doibase
  10.1021/jacs.6b05111} {\bibfield  {journal} {\bibinfo  {journal} {J. Am.
  Chem. Soc.}\ }\textbf {\bibinfo {volume} {138}},\ \bibinfo {pages}
  {11211--11218} (\bibinfo {year} {2016}{\natexlab{b}})}\BibitemShut {NoStop}%
\bibitem [{\citenamefont {Tang}\ \emph {et~al.}(2017)\citenamefont {Tang},
  \citenamefont {Dalzini}, \citenamefont {Li}, \citenamefont {Feng},
  \citenamefont {Chien}, \citenamefont {Song},\ and\ \citenamefont
  {Hu}}]{Tang2017JPCL}%
  \BibitemOpen
  \bibfield  {author} {\bibinfo {author} {\bibfnamefont {M.}~\bibnamefont
  {Tang}}, \bibinfo {author} {\bibfnamefont {A.}~\bibnamefont {Dalzini}},
  \bibinfo {author} {\bibfnamefont {X.}~\bibnamefont {Li}}, \bibinfo {author}
  {\bibfnamefont {X.}~\bibnamefont {Feng}}, \bibinfo {author} {\bibfnamefont
  {P.-H.}\ \bibnamefont {Chien}}, \bibinfo {author} {\bibfnamefont
  {L.}~\bibnamefont {Song}}, \ and\ \bibinfo {author} {\bibfnamefont {Y.-Y.}\
  \bibnamefont {Hu}},\ }\bibfield  {title} {\enquote {\bibinfo {title}
  {{Operando EPR for Simultaneous Monitoring of Anionic and Cationic Redox
  Processes in Li-Rich Metal Oxide Cathodes}},}\ }\href {\doibase
  10.1021/acs.jpclett.7b01425} {\bibfield  {journal} {\bibinfo  {journal} {J.
  Phys. Chem. Lett.}\ }\textbf {\bibinfo {volume} {8}},\ \bibinfo {pages}
  {4009--4016} (\bibinfo {year} {2017})}\BibitemShut {NoStop}%
\bibitem [{\citenamefont {Lu}\ \emph {et~al.}(2001)\citenamefont {Lu},
  \citenamefont {MacNeil},\ and\ \citenamefont {Dahn}}]{Lu2001}%
  \BibitemOpen
  \bibfield  {author} {\bibinfo {author} {\bibfnamefont {Z.}~\bibnamefont
  {Lu}}, \bibinfo {author} {\bibfnamefont {D.~D.}\ \bibnamefont {MacNeil}}, \
  and\ \bibinfo {author} {\bibfnamefont {J.~R.}\ \bibnamefont {Dahn}},\
  }\bibfield  {title} {\enquote {\bibinfo {title} {{Layered Cathode Materials
  Li[Ni$_x$Li$_{1/3-2x/3}$Mn$_{2/3-x/3}$]O$_2$ for Lithium-Ion Batteries}},}\
  }\href {\doibase 10.1149/1.1407994} {\bibfield  {journal} {\bibinfo
  {journal} {Electrochem. Solid-State Lett.}\ }\textbf {\bibinfo {volume}
  {4}},\ \bibinfo {pages} {A191--A194} (\bibinfo {year} {2001})}\BibitemShut
  {NoStop}%
\bibitem [{\citenamefont {Matsunaga}\ \emph {et~al.}(2016)\citenamefont
  {Matsunaga}, \citenamefont {Komatsu}, \citenamefont {Shimoda}, \citenamefont
  {Minato}, \citenamefont {Yonemura}, \citenamefont {Kamiyama}, \citenamefont
  {Kobayashi}, \citenamefont {Kato}, \citenamefont {Hirayama}, \citenamefont
  {Ikuhara}, \citenamefont {Arai}, \citenamefont {Ukyo}, \citenamefont
  {Uchimoto},\ and\ \citenamefont {Ogumi}}]{Matsunaga2016JPCL}%
  \BibitemOpen
  \bibfield  {author} {\bibinfo {author} {\bibfnamefont {T.}~\bibnamefont
  {Matsunaga}}, \bibinfo {author} {\bibfnamefont {H.}~\bibnamefont {Komatsu}},
  \bibinfo {author} {\bibfnamefont {K.}~\bibnamefont {Shimoda}}, \bibinfo
  {author} {\bibfnamefont {T.}~\bibnamefont {Minato}}, \bibinfo {author}
  {\bibfnamefont {M.}~\bibnamefont {Yonemura}}, \bibinfo {author}
  {\bibfnamefont {T.}~\bibnamefont {Kamiyama}}, \bibinfo {author}
  {\bibfnamefont {S.}~\bibnamefont {Kobayashi}}, \bibinfo {author}
  {\bibfnamefont {T.}~\bibnamefont {Kato}}, \bibinfo {author} {\bibfnamefont
  {T.}~\bibnamefont {Hirayama}}, \bibinfo {author} {\bibfnamefont
  {Y.}~\bibnamefont {Ikuhara}}, \bibinfo {author} {\bibfnamefont
  {H.}~\bibnamefont {Arai}}, \bibinfo {author} {\bibfnamefont {Y.}~\bibnamefont
  {Ukyo}}, \bibinfo {author} {\bibfnamefont {Y.}~\bibnamefont {Uchimoto}}, \
  and\ \bibinfo {author} {\bibfnamefont {Z.}~\bibnamefont {Ogumi}},\ }\bibfield
   {title} {\enquote {\bibinfo {title} {{Structural Understanding of Superior
  Battery Properties of Partially Ni-Doped \ce{Li2MnO3} as Cathode
  Material}},}\ }\href {\doibase 10.1021/acs.jpclett.6b00587} {\bibfield
  {journal} {\bibinfo  {journal} {J. Phys. Chem. Lett.}\ }\textbf {\bibinfo
  {volume} {7}},\ \bibinfo {pages} {2063--2067} (\bibinfo {year}
  {2016})}\BibitemShut {NoStop}%
\bibitem [{\citenamefont {Ma}\ \emph {et~al.}(2014)\citenamefont {Ma},
  \citenamefont {Zhou}, \citenamefont {Gao}, \citenamefont {Kong},
  \citenamefont {Wang}, \citenamefont {Yang},\ and\ \citenamefont
  {Chen}}]{Ma2014CEJ}%
  \BibitemOpen
  \bibfield  {author} {\bibinfo {author} {\bibfnamefont {J.}~\bibnamefont
  {Ma}}, \bibinfo {author} {\bibfnamefont {Y.-N.}\ \bibnamefont {Zhou}},
  \bibinfo {author} {\bibfnamefont {Y.}~\bibnamefont {Gao}}, \bibinfo {author}
  {\bibfnamefont {Q.}~\bibnamefont {Kong}}, \bibinfo {author} {\bibfnamefont
  {Z.}~\bibnamefont {Wang}}, \bibinfo {author} {\bibfnamefont {X.-Q.}\
  \bibnamefont {Yang}}, \ and\ \bibinfo {author} {\bibfnamefont
  {L.}~\bibnamefont {Chen}},\ }\bibfield  {title} {\enquote {\bibinfo {title}
  {{Molybdenum Substitution for Improving the Charge Compensation and Activity
  of \ce{Li2MnO3}}},}\ }\href {\doibase 10.1002/chem.201402727} {\bibfield
  {journal} {\bibinfo  {journal} {Chem. Eur. J.}\ }\textbf {\bibinfo {volume}
  {20}},\ \bibinfo {pages} {8723--8730} (\bibinfo {year} {2014})}\BibitemShut
  {NoStop}%
\bibitem [{\citenamefont {Mori}\ \emph {et~al.}(2011)\citenamefont {Mori},
  \citenamefont {Sakaebe}, \citenamefont {Shikano}, \citenamefont {Kojitani},
  \citenamefont {Tatsumi},\ and\ \citenamefont {Inaguma}}]{Mori2011JPS}%
  \BibitemOpen
  \bibfield  {author} {\bibinfo {author} {\bibfnamefont {D.}~\bibnamefont
  {Mori}}, \bibinfo {author} {\bibfnamefont {H.}~\bibnamefont {Sakaebe}},
  \bibinfo {author} {\bibfnamefont {M.}~\bibnamefont {Shikano}}, \bibinfo
  {author} {\bibfnamefont {H.}~\bibnamefont {Kojitani}}, \bibinfo {author}
  {\bibfnamefont {K.}~\bibnamefont {Tatsumi}}, \ and\ \bibinfo {author}
  {\bibfnamefont {Y.}~\bibnamefont {Inaguma}},\ }\bibfield  {title} {\enquote
  {\bibinfo {title} {{Synthesis, phase relation and electrical and
  electrochemical properties of ruthenium-substituted \ce{Li2MnO3} as a novel
  cathode material}},}\ }\href {\doibase
  http://dx.doi.org/10.1016/j.jpowsour.2010.11.150} {\bibfield  {journal}
  {\bibinfo  {journal} {J. Power Sources}\ }\textbf {\bibinfo {volume} {196}},\
  \bibinfo {pages} {6934--6938} (\bibinfo {year} {2011})}\BibitemShut {NoStop}%
\bibitem [{\citenamefont {Kong}\ \emph
  {et~al.}(2015{\natexlab{a}})\citenamefont {Kong}, \citenamefont {Longo},
  \citenamefont {Park}, \citenamefont {Yoon}, \citenamefont {Yeon},
  \citenamefont {Park}, \citenamefont {Wang}, \citenamefont {KC}, \citenamefont
  {Doo},\ and\ \citenamefont {Cho}}]{Kong2015JMCA}%
  \BibitemOpen
  \bibfield  {author} {\bibinfo {author} {\bibfnamefont {F.}~\bibnamefont
  {Kong}}, \bibinfo {author} {\bibfnamefont {R.~C.}\ \bibnamefont {Longo}},
  \bibinfo {author} {\bibfnamefont {M.-S.}\ \bibnamefont {Park}}, \bibinfo
  {author} {\bibfnamefont {J.}~\bibnamefont {Yoon}}, \bibinfo {author}
  {\bibfnamefont {D.-H.}\ \bibnamefont {Yeon}}, \bibinfo {author}
  {\bibfnamefont {J.-H.}\ \bibnamefont {Park}}, \bibinfo {author}
  {\bibfnamefont {W.-H.}\ \bibnamefont {Wang}}, \bibinfo {author}
  {\bibfnamefont {S.}~\bibnamefont {KC}}, \bibinfo {author} {\bibfnamefont
  {S.-G.}\ \bibnamefont {Doo}}, \ and\ \bibinfo {author} {\bibfnamefont
  {K.}~\bibnamefont {Cho}},\ }\bibfield  {title} {\enquote {\bibinfo {title}
  {{{\it Ab initio} study of doping effects on \ce{LiMnO2} and \ce{Li2MnO3}
  cathode materials for Li-ion batteries}},}\ }\href {\doibase
  10.1039/C5TA01445J} {\bibfield  {journal} {\bibinfo  {journal} {J. Mater.
  Chem. A}\ }\textbf {\bibinfo {volume} {3}},\ \bibinfo {pages} {8489--8500}
  (\bibinfo {year} {2015}{\natexlab{a}})}\BibitemShut {NoStop}%
\bibitem [{\citenamefont {Kong}\ \emph
  {et~al.}(2015{\natexlab{b}})\citenamefont {Kong}, \citenamefont {Longo},
  \citenamefont {Yeon}, \citenamefont {Yoon}, \citenamefont {Park},
  \citenamefont {Liang}, \citenamefont {KC}, \citenamefont {Zheng},
  \citenamefont {Doo},\ and\ \citenamefont {Cho}}]{Kong2015JPCC}%
  \BibitemOpen
  \bibfield  {author} {\bibinfo {author} {\bibfnamefont {F.}~\bibnamefont
  {Kong}}, \bibinfo {author} {\bibfnamefont {R.~C.}\ \bibnamefont {Longo}},
  \bibinfo {author} {\bibfnamefont {D.-H.}\ \bibnamefont {Yeon}}, \bibinfo
  {author} {\bibfnamefont {J.}~\bibnamefont {Yoon}}, \bibinfo {author}
  {\bibfnamefont {J.-H.}\ \bibnamefont {Park}}, \bibinfo {author}
  {\bibfnamefont {C.}~\bibnamefont {Liang}}, \bibinfo {author} {\bibfnamefont
  {S.}~\bibnamefont {KC}}, \bibinfo {author} {\bibfnamefont {Y.}~\bibnamefont
  {Zheng}}, \bibinfo {author} {\bibfnamefont {S.-G.}\ \bibnamefont {Doo}}, \
  and\ \bibinfo {author} {\bibfnamefont {K.}~\bibnamefont {Cho}},\ }\bibfield
  {title} {\enquote {\bibinfo {title} {{Multivalent Li-Site Doping of Mn Oxides
  for Li-Ion Batteries}},}\ }\href {\doibase 10.1021/acs.jpcc.5b06844}
  {\bibfield  {journal} {\bibinfo  {journal} {J. Phys. Chem. C}\ }\textbf
  {\bibinfo {volume} {119}},\ \bibinfo {pages} {21904--21912} (\bibinfo {year}
  {2015}{\natexlab{b}})}\BibitemShut {NoStop}%
\bibitem [{\citenamefont {Lee}\ \emph {et~al.}(2015)\citenamefont {Lee},
  \citenamefont {Moon}, \citenamefont {Lee}, \citenamefont {Yu}, \citenamefont
  {Kim},\ and\ \citenamefont {Park}}]{Lee2015JPS}%
  \BibitemOpen
  \bibfield  {author} {\bibinfo {author} {\bibfnamefont {S.~H.}\ \bibnamefont
  {Lee}}, \bibinfo {author} {\bibfnamefont {J.-S.}\ \bibnamefont {Moon}},
  \bibinfo {author} {\bibfnamefont {M.-S.}\ \bibnamefont {Lee}}, \bibinfo
  {author} {\bibfnamefont {T.-H.}\ \bibnamefont {Yu}}, \bibinfo {author}
  {\bibfnamefont {H.}~\bibnamefont {Kim}}, \ and\ \bibinfo {author}
  {\bibfnamefont {B.~M.}\ \bibnamefont {Park}},\ }\bibfield  {title} {\enquote
  {\bibinfo {title} {{Enhancing phase stability and kinetics of lithium-rich
  layered oxide for an ultra-high performing cathode in Li-ion batteries}},}\
  }\href {\doibase http://dx.doi.org/10.1016/j.jpowsour.2015.01.158} {\bibfield
   {journal} {\bibinfo  {journal} {J. Power Sources}\ }\textbf {\bibinfo
  {volume} {281}},\ \bibinfo {pages} {77--84} (\bibinfo {year}
  {2015})}\BibitemShut {NoStop}%
\bibitem [{\citenamefont {Gao}\ \emph {et~al.}(2014)\citenamefont {Gao},
  \citenamefont {Ma}, \citenamefont {Wang}, \citenamefont {Lu}, \citenamefont
  {Bai}, \citenamefont {Wang},\ and\ \citenamefont {Chen}}]{Gao2014JMCA}%
  \BibitemOpen
  \bibfield  {author} {\bibinfo {author} {\bibfnamefont {Y.}~\bibnamefont
  {Gao}}, \bibinfo {author} {\bibfnamefont {J.}~\bibnamefont {Ma}}, \bibinfo
  {author} {\bibfnamefont {X.}~\bibnamefont {Wang}}, \bibinfo {author}
  {\bibfnamefont {X.}~\bibnamefont {Lu}}, \bibinfo {author} {\bibfnamefont
  {Y.}~\bibnamefont {Bai}}, \bibinfo {author} {\bibfnamefont {Z.}~\bibnamefont
  {Wang}}, \ and\ \bibinfo {author} {\bibfnamefont {L.}~\bibnamefont {Chen}},\
  }\bibfield  {title} {\enquote {\bibinfo {title} {{Improved electron/Li-ion
  transport and oxygen stability of Mo-doped \ce{Li2MnO3}}},}\ }\href {\doibase
  10.1039/C3TA15236G} {\bibfield  {journal} {\bibinfo  {journal} {J. Mater.
  Chem. A}\ }\textbf {\bibinfo {volume} {2}},\ \bibinfo {pages} {4811--4818}
  (\bibinfo {year} {2014})}\BibitemShut {NoStop}%
\bibitem [{\citenamefont {Gao}\ \emph {et~al.}(2015)\citenamefont {Gao},
  \citenamefont {Wang}, \citenamefont {Ma}, \citenamefont {Wang},\ and\
  \citenamefont {Chen}}]{Gao2015CM}%
  \BibitemOpen
  \bibfield  {author} {\bibinfo {author} {\bibfnamefont {Y.}~\bibnamefont
  {Gao}}, \bibinfo {author} {\bibfnamefont {X.}~\bibnamefont {Wang}}, \bibinfo
  {author} {\bibfnamefont {J.}~\bibnamefont {Ma}}, \bibinfo {author}
  {\bibfnamefont {Z.}~\bibnamefont {Wang}}, \ and\ \bibinfo {author}
  {\bibfnamefont {L.}~\bibnamefont {Chen}},\ }\bibfield  {title} {\enquote
  {\bibinfo {title} {{Selecting Substituent Elements for Li-Rich Mn-Based
  Cathode Materials by Density Functional Theory (DFT) Calculations}},}\ }\href
  {\doibase 10.1021/acs.chemmater.5b00875} {\bibfield  {journal} {\bibinfo
  {journal} {Chem. Mater.}\ }\textbf {\bibinfo {volume} {27}},\ \bibinfo
  {pages} {3456--3461} (\bibinfo {year} {2015})}\BibitemShut {NoStop}%
\bibitem [{\citenamefont {Yang}\ \emph {et~al.}(2017)\citenamefont {Yang},
  \citenamefont {Kim}, \citenamefont {Kim}, \citenamefont {Cho}, \citenamefont
  {Choi},\ and\ \citenamefont {Nam}}]{Yang2017AFM}%
  \BibitemOpen
  \bibfield  {author} {\bibinfo {author} {\bibfnamefont {M.~Y.}\ \bibnamefont
  {Yang}}, \bibinfo {author} {\bibfnamefont {S.}~\bibnamefont {Kim}}, \bibinfo
  {author} {\bibfnamefont {K.}~\bibnamefont {Kim}}, \bibinfo {author}
  {\bibfnamefont {W.}~\bibnamefont {Cho}}, \bibinfo {author} {\bibfnamefont
  {J.~W.}\ \bibnamefont {Choi}}, \ and\ \bibinfo {author} {\bibfnamefont
  {Y.~S.}\ \bibnamefont {Nam}},\ }\bibfield  {title} {\enquote {\bibinfo
  {title} {{Role of Ordered Ni Atoms in Li Layers for Li-Rich Layered Cathode
  Materials}},}\ }\href {\doibase 10.1002/adfm.201700982} {\bibfield  {journal}
  {\bibinfo  {journal} {Adv. Funct. Mater.}\ }\textbf {\bibinfo {volume}
  {27}},\ \bibinfo {pages} {1700982} (\bibinfo {year} {2017})}\BibitemShut
  {NoStop}%
\bibitem [{\citenamefont {Anisimov}\ \emph {et~al.}(1997)\citenamefont
  {Anisimov}, \citenamefont {Aryasetiawan},\ and\ \citenamefont
  {Lichtenstein}}]{Anisimov1997JPCM}%
  \BibitemOpen
  \bibfield  {author} {\bibinfo {author} {\bibfnamefont {V.~I.}\ \bibnamefont
  {Anisimov}}, \bibinfo {author} {\bibfnamefont {F.}~\bibnamefont
  {Aryasetiawan}}, \ and\ \bibinfo {author} {\bibfnamefont {A.~I.}\
  \bibnamefont {Lichtenstein}},\ }\bibfield  {title} {\enquote {\bibinfo
  {title} {{First-principles calculations of the electronic structure and
  spectra of strongly correlated systems: the LDA$+U$ method}},}\ }\href
  {http://stacks.iop.org/0953-8984/9/i=4/a=002} {\bibfield  {journal} {\bibinfo
   {journal} {J. Phys.: Condens. Matter}\ }\textbf {\bibinfo {volume} {9}},\
  \bibinfo {pages} {767} (\bibinfo {year} {1997})}\BibitemShut {NoStop}%
\bibitem [{\citenamefont {Koyama}\ \emph {et~al.}(2009)\citenamefont {Koyama},
  \citenamefont {Tanaka}, \citenamefont {Nagao},\ and\ \citenamefont
  {Kanno}}]{Koyama2009}%
  \BibitemOpen
  \bibfield  {author} {\bibinfo {author} {\bibfnamefont {Y.}~\bibnamefont
  {Koyama}}, \bibinfo {author} {\bibfnamefont {I.}~\bibnamefont {Tanaka}},
  \bibinfo {author} {\bibfnamefont {M.}~\bibnamefont {Nagao}}, \ and\ \bibinfo
  {author} {\bibfnamefont {R.}~\bibnamefont {Kanno}},\ }\bibfield  {title}
  {\enquote {\bibinfo {title} {{First-principles study on lithium removal from
  \ce{Li2MnO3}}},}\ }\href {\doibase
  http://dx.doi.org/10.1016/j.jpowsour.2008.07.073} {\bibfield  {journal}
  {\bibinfo  {journal} {J. Power Sources}\ }\textbf {\bibinfo {volume} {189}},\
  \bibinfo {pages} {798--801} (\bibinfo {year} {2009})}\BibitemShut {NoStop}%
\bibitem [{\citenamefont {Xiao}\ \emph
  {et~al.}(2012{\natexlab{a}})\citenamefont {Xiao}, \citenamefont {Li},\ and\
  \citenamefont {Chen}}]{Xiao2012}%
  \BibitemOpen
  \bibfield  {author} {\bibinfo {author} {\bibfnamefont {R.}~\bibnamefont
  {Xiao}}, \bibinfo {author} {\bibfnamefont {H.}~\bibnamefont {Li}}, \ and\
  \bibinfo {author} {\bibfnamefont {L.}~\bibnamefont {Chen}},\ }\bibfield
  {title} {\enquote {\bibinfo {title} {{Density Functional Investigation on
  \ce{Li2MnO3}}},}\ }\href {\doibase 10.1021/cm3027219} {\bibfield  {journal}
  {\bibinfo  {journal} {Chem. Mater.}\ }\textbf {\bibinfo {volume} {24}},\
  \bibinfo {pages} {4242--4251} (\bibinfo {year}
  {2012}{\natexlab{a}})}\BibitemShut {NoStop}%
\bibitem [{\citenamefont {Xiao}\ \emph
  {et~al.}(2012{\natexlab{b}})\citenamefont {Xiao}, \citenamefont {Deng},
  \citenamefont {Manthiram},\ and\ \citenamefont {Henkelman}}]{Xiao2012JPCC}%
  \BibitemOpen
  \bibfield  {author} {\bibinfo {author} {\bibfnamefont {P.}~\bibnamefont
  {Xiao}}, \bibinfo {author} {\bibfnamefont {Z.~Q.}\ \bibnamefont {Deng}},
  \bibinfo {author} {\bibfnamefont {A.}~\bibnamefont {Manthiram}}, \ and\
  \bibinfo {author} {\bibfnamefont {G.}~\bibnamefont {Henkelman}},\ }\bibfield
  {title} {\enquote {\bibinfo {title} {{Calculations of Oxygen Stability in
  Lithium-Rich Layered Cathodes}},}\ }\href {\doibase 10.1021/jp3058788}
  {\bibfield  {journal} {\bibinfo  {journal} {J. Phys. Chem. C}\ }\textbf
  {\bibinfo {volume} {116}},\ \bibinfo {pages} {23201--23204} (\bibinfo {year}
  {2012}{\natexlab{b}})}\BibitemShut {NoStop}%
\bibitem [{\citenamefont {Marusczyk}\ \emph {et~al.}(2017)\citenamefont
  {Marusczyk}, \citenamefont {Albina}, \citenamefont {Hammerschmidt},
  \citenamefont {Drautz}, \citenamefont {Eckl},\ and\ \citenamefont
  {Henkelman}}]{Marusczyk2017JMCA}%
  \BibitemOpen
  \bibfield  {author} {\bibinfo {author} {\bibfnamefont {A.}~\bibnamefont
  {Marusczyk}}, \bibinfo {author} {\bibfnamefont {J.-M.}\ \bibnamefont
  {Albina}}, \bibinfo {author} {\bibfnamefont {T.}~\bibnamefont
  {Hammerschmidt}}, \bibinfo {author} {\bibfnamefont {R.}~\bibnamefont
  {Drautz}}, \bibinfo {author} {\bibfnamefont {T.}~\bibnamefont {Eckl}}, \ and\
  \bibinfo {author} {\bibfnamefont {G.}~\bibnamefont {Henkelman}},\ }\bibfield
  {title} {\enquote {\bibinfo {title} {{Oxygen activity and peroxide formation
  as charge compensation mechanisms in \ce{Li2MnO3}}},}\ }\href {\doibase
  10.1039/C7TA04164K} {\bibfield  {journal} {\bibinfo  {journal} {J. Mater.
  Chem. A}\ }\textbf {\bibinfo {volume} {5}},\ \bibinfo {pages} {15183--15190}
  (\bibinfo {year} {2017})}\BibitemShut {NoStop}%
\bibitem [{\citenamefont {Heyd}\ \emph {et~al.}(2003)\citenamefont {Heyd},
  \citenamefont {Scuseria},\ and\ \citenamefont {Ernzerhof}}]{heyd:8207}%
  \BibitemOpen
  \bibfield  {author} {\bibinfo {author} {\bibfnamefont {J.}~\bibnamefont
  {Heyd}}, \bibinfo {author} {\bibfnamefont {G.~E.}\ \bibnamefont {Scuseria}},
  \ and\ \bibinfo {author} {\bibfnamefont {M.}~\bibnamefont {Ernzerhof}},\
  }\bibfield  {title} {\enquote {\bibinfo {title} {{Hybrid functionals based on
  a screened Coulomb potential}},}\ }\href {\doibase 10.1063/1.1564060}
  {\bibfield  {journal} {\bibinfo  {journal} {J. Chem. Phys.}\ }\textbf
  {\bibinfo {volume} {118}},\ \bibinfo {pages} {8207--8215} (\bibinfo {year}
  {2003})}\BibitemShut {NoStop}%
\bibitem [{\citenamefont {Bl\"ochl}(1994)}]{PAW1}%
  \BibitemOpen
  \bibfield  {author} {\bibinfo {author} {\bibfnamefont {P.~E.}\ \bibnamefont
  {Bl\"ochl}},\ }\bibfield  {title} {\enquote {\bibinfo {title} {Projector
  augmented-wave method},}\ }\href {\doibase 10.1103/PhysRevB.50.17953}
  {\bibfield  {journal} {\bibinfo  {journal} {Phys. Rev. B}\ }\textbf {\bibinfo
  {volume} {50}},\ \bibinfo {pages} {17953--17979} (\bibinfo {year}
  {1994})}\BibitemShut {NoStop}%
\bibitem [{\citenamefont {Kresse}\ and\ \citenamefont
  {Furthm\"uller}(1996)}]{VASP2}%
  \BibitemOpen
  \bibfield  {author} {\bibinfo {author} {\bibfnamefont {G.}~\bibnamefont
  {Kresse}}\ and\ \bibinfo {author} {\bibfnamefont {J.}~\bibnamefont
  {Furthm\"uller}},\ }\bibfield  {title} {\enquote {\bibinfo {title} {Efficient
  iterative schemes for ab initio total-energy calculations using a plane-wave
  basis set},}\ }\href {\doibase 10.1103/PhysRevB.54.11169} {\bibfield
  {journal} {\bibinfo  {journal} {Phys. Rev. B}\ }\textbf {\bibinfo {volume}
  {54}},\ \bibinfo {pages} {11169--11186} (\bibinfo {year} {1996})}\BibitemShut
  {NoStop}%
\bibitem [{\citenamefont {Hoang}\ and\ \citenamefont
  {Johannes}(2014)}]{Hoang2014JMCA}%
  \BibitemOpen
  \bibfield  {author} {\bibinfo {author} {\bibfnamefont {K.}~\bibnamefont
  {Hoang}}\ and\ \bibinfo {author} {\bibfnamefont {M.~D.}\ \bibnamefont
  {Johannes}},\ }\bibfield  {title} {\enquote {\bibinfo {title} {Defect
  chemistry in layered transition-metal oxides from screened hybrid density
  functional calculations},}\ }\href {\doibase 10.1039/C4TA00673A} {\bibfield
  {journal} {\bibinfo  {journal} {J. Mater. Chem. A}\ }\textbf {\bibinfo
  {volume} {2}},\ \bibinfo {pages} {5224--5235} (\bibinfo {year}
  {2014})}\BibitemShut {NoStop}%
\bibitem [{\citenamefont {Freysoldt}\ \emph {et~al.}(2009)\citenamefont
  {Freysoldt}, \citenamefont {Neugebauer},\ and\ \citenamefont {{Van de
  Walle}}}]{Freysoldt2009}%
  \BibitemOpen
  \bibfield  {author} {\bibinfo {author} {\bibfnamefont {C.}~\bibnamefont
  {Freysoldt}}, \bibinfo {author} {\bibfnamefont {J.}~\bibnamefont
  {Neugebauer}}, \ and\ \bibinfo {author} {\bibfnamefont {C.~G.}\ \bibnamefont
  {{Van de Walle}}},\ }\bibfield  {title} {\enquote {\bibinfo {title} {{Fully
  \textit{Ab Initio} Finite-Size Corrections for Charged-Defect Supercell
  Calculations}},}\ }\href {\doibase 10.1103/PhysRevLett.102.016402} {\bibfield
   {journal} {\bibinfo  {journal} {Phys. Rev. Lett.}\ }\textbf {\bibinfo
  {volume} {102}},\ \bibinfo {pages} {016402} (\bibinfo {year}
  {2009})}\BibitemShut {NoStop}%
\bibitem [{\citenamefont {{Van de Walle}}\ and\ \citenamefont
  {Neugebauer}(2004)}]{walle:3851}%
  \BibitemOpen
  \bibfield  {author} {\bibinfo {author} {\bibfnamefont {C.~G.}\ \bibnamefont
  {{Van de Walle}}}\ and\ \bibinfo {author} {\bibfnamefont {J.}~\bibnamefont
  {Neugebauer}},\ }\bibfield  {title} {\enquote {\bibinfo {title}
  {{First-principles calculations for defects and impurities: Applications to
  III-nitrides}},}\ }\href {\doibase 10.1063/1.1682673} {\bibfield  {journal}
  {\bibinfo  {journal} {J. Appl. Phys.}\ }\textbf {\bibinfo {volume} {95}},\
  \bibinfo {pages} {3851--3879} (\bibinfo {year} {2004})}\BibitemShut {NoStop}%
\bibitem [{\citenamefont {Cupid}\ \emph {et~al.}(2016)\citenamefont {Cupid},
  \citenamefont {Li}, \citenamefont {Gebert}, \citenamefont {Reif},
  \citenamefont {Flandorfer},\ and\ \citenamefont {Seifert}}]{Cupid2016JCSJ}%
  \BibitemOpen
  \bibfield  {author} {\bibinfo {author} {\bibfnamefont {D.~M.}\ \bibnamefont
  {Cupid}}, \bibinfo {author} {\bibfnamefont {D.}~\bibnamefont {Li}}, \bibinfo
  {author} {\bibfnamefont {C.}~\bibnamefont {Gebert}}, \bibinfo {author}
  {\bibfnamefont {A.}~\bibnamefont {Reif}}, \bibinfo {author} {\bibfnamefont
  {H.}~\bibnamefont {Flandorfer}}, \ and\ \bibinfo {author} {\bibfnamefont
  {H.~J.}\ \bibnamefont {Seifert}},\ }\bibfield  {title} {\enquote {\bibinfo
  {title} {{Enthalpy of formation and heat capacity of \ce{Li2MnO3}}},}\ }\href
  {\doibase 10.2109/jcersj2.16116} {\bibfield  {journal} {\bibinfo  {journal}
  {J. Ceram. Soc. Japan}\ }\textbf {\bibinfo {volume} {124}},\ \bibinfo {pages}
  {1072--1082} (\bibinfo {year} {2016})}\BibitemShut {NoStop}%
\bibitem [{\citenamefont {Robertson}\ and\ \citenamefont
  {Bruce}(2002)}]{Robertson2002}%
  \BibitemOpen
  \bibfield  {author} {\bibinfo {author} {\bibfnamefont {A.~D.}\ \bibnamefont
  {Robertson}}\ and\ \bibinfo {author} {\bibfnamefont {P.~G.}\ \bibnamefont
  {Bruce}},\ }\bibfield  {title} {\enquote {\bibinfo {title} {{The origin of
  electrochemical activity in \ce{Li2MnO3}}},}\ }\href {\doibase
  10.1039/B207945C} {\bibfield  {journal} {\bibinfo  {journal} {Chem. Commun.}\
  ,\ \bibinfo {pages} {2790--2791}} (\bibinfo {year} {2002})}\BibitemShut
  {NoStop}%
\bibitem [{\citenamefont {Kubota}\ \emph {et~al.}(2012)\citenamefont {Kubota},
  \citenamefont {Kaneko}, \citenamefont {Hirayama}, \citenamefont {Yonemura},
  \citenamefont {Imanari}, \citenamefont {Nakane},\ and\ \citenamefont
  {Kanno}}]{Kubota2012}%
  \BibitemOpen
  \bibfield  {author} {\bibinfo {author} {\bibfnamefont {K.}~\bibnamefont
  {Kubota}}, \bibinfo {author} {\bibfnamefont {T.}~\bibnamefont {Kaneko}},
  \bibinfo {author} {\bibfnamefont {M.}~\bibnamefont {Hirayama}}, \bibinfo
  {author} {\bibfnamefont {M.}~\bibnamefont {Yonemura}}, \bibinfo {author}
  {\bibfnamefont {Y.}~\bibnamefont {Imanari}}, \bibinfo {author} {\bibfnamefont
  {K.}~\bibnamefont {Nakane}}, \ and\ \bibinfo {author} {\bibfnamefont
  {R.}~\bibnamefont {Kanno}},\ }\bibfield  {title} {\enquote {\bibinfo {title}
  {{Direct synthesis of oxygen-deficient Li$_2$MnO$_{3-x}$ for high capacity
  lithium battery electrodes}},}\ }\href {\doibase
  http://dx.doi.org/10.1016/j.jpowsour.2012.05.061} {\bibfield  {journal}
  {\bibinfo  {journal} {J. Power Sources}\ }\textbf {\bibinfo {volume} {216}},\
  \bibinfo {pages} {249--255} (\bibinfo {year} {2012})}\BibitemShut {NoStop}%
\bibitem [{\citenamefont {Aydinol}\ \emph {et~al.}(1997)\citenamefont
  {Aydinol}, \citenamefont {Kohan}, \citenamefont {Ceder}, \citenamefont
  {Cho},\ and\ \citenamefont {Joannopoulos}}]{Aydinol1997}%
  \BibitemOpen
  \bibfield  {author} {\bibinfo {author} {\bibfnamefont {M.~K.}\ \bibnamefont
  {Aydinol}}, \bibinfo {author} {\bibfnamefont {A.~F.}\ \bibnamefont {Kohan}},
  \bibinfo {author} {\bibfnamefont {G.}~\bibnamefont {Ceder}}, \bibinfo
  {author} {\bibfnamefont {K.}~\bibnamefont {Cho}}, \ and\ \bibinfo {author}
  {\bibfnamefont {J.}~\bibnamefont {Joannopoulos}},\ }\bibfield  {title}
  {\enquote {\bibinfo {title} {Ab initio study of lithium intercalation in
  metal oxides and metal dichalcogenides},}\ }\href {\doibase
  10.1103/PhysRevB.56.1354} {\bibfield  {journal} {\bibinfo  {journal} {Phys.
  Rev. B}\ }\textbf {\bibinfo {volume} {56}},\ \bibinfo {pages} {1354--1365}
  (\bibinfo {year} {1997})}\BibitemShut {NoStop}%
\bibitem [{\citenamefont {Seo}\ \emph {et~al.}(2016)\citenamefont {Seo},
  \citenamefont {Lee}, \citenamefont {Urban}, \citenamefont {Malik},
  \citenamefont {Kang},\ and\ \citenamefont {Ceder}}]{Seo2016NC}%
  \BibitemOpen
  \bibfield  {author} {\bibinfo {author} {\bibfnamefont {D.-H.}\ \bibnamefont
  {Seo}}, \bibinfo {author} {\bibfnamefont {J.}~\bibnamefont {Lee}}, \bibinfo
  {author} {\bibfnamefont {A.}~\bibnamefont {Urban}}, \bibinfo {author}
  {\bibfnamefont {R.}~\bibnamefont {Malik}}, \bibinfo {author} {\bibfnamefont
  {S.}~\bibnamefont {Kang}}, \ and\ \bibinfo {author} {\bibfnamefont
  {G.}~\bibnamefont {Ceder}},\ }\bibfield  {title} {\enquote {\bibinfo {title}
  {{The structural and chemical origin of the oxygen redox activity in layered
  and cation-disordered Li-excess cathode materials}},}\ }\href {\doibase
  10.1038/nchem.2524} {\bibfield  {journal} {\bibinfo  {journal} {Nature
  Chem.}\ }\textbf {\bibinfo {volume} {8}},\ \bibinfo {pages} {692--697}
  (\bibinfo {year} {2016})}\BibitemShut {NoStop}%
\bibitem [{SM()}]{SM}%
  \BibitemOpen
  \href@noop {} {}\bibinfo {note} {See Supplemental Material at [URL will be
  inserted by publisher] for the electronic structures of Li$_{2-x}$MnO$_3$
  ($0\leq x\leq 2$) calculated using the HSE06 hybrid functional with two
  different values for the Hartree-Fock mixing parameter, $\alpha=25\%$ and
  $17\%$.}\BibitemShut {Stop}%
\bibitem [{\citenamefont {Chen}\ and\ \citenamefont
  {Islam}(2016)}]{Chen2016CM}%
  \BibitemOpen
  \bibfield  {author} {\bibinfo {author} {\bibfnamefont {H.}~\bibnamefont
  {Chen}}\ and\ \bibinfo {author} {\bibfnamefont {M.~S.}\ \bibnamefont
  {Islam}},\ }\bibfield  {title} {\enquote {\bibinfo {title} {{Lithium
  Extraction Mechanism in Li-Rich \ce{Li2MnO3} Involving Oxygen Hole Formation
  and Dimerization}},}\ }\href {\doibase 10.1021/acs.chemmater.6b02870}
  {\bibfield  {journal} {\bibinfo  {journal} {Chem. Mater.}\ }\textbf {\bibinfo
  {volume} {28}},\ \bibinfo {pages} {6656--6663} (\bibinfo {year}
  {2016})}\BibitemShut {NoStop}%
\bibitem [{\citenamefont {Hoang}\ and\ \citenamefont
  {Johannes}(2012)}]{Hoang2012JPS}%
  \BibitemOpen
  \bibfield  {author} {\bibinfo {author} {\bibfnamefont {K.}~\bibnamefont
  {Hoang}}\ and\ \bibinfo {author} {\bibfnamefont {M.~D.}\ \bibnamefont
  {Johannes}},\ }\bibfield  {title} {\enquote {\bibinfo {title}
  {{First-principles studies of the effects of impurities on the ionic and
  electronic conduction in \ce{LiFePO4}}},}\ }\href {\doibase
  http://dx.doi.org/10.1016/j.jpowsour.2012.01.126} {\bibfield  {journal}
  {\bibinfo  {journal} {J. Power Sources}\ }\textbf {\bibinfo {volume} {206}},\
  \bibinfo {pages} {274--281} (\bibinfo {year} {2012})}\BibitemShut {NoStop}%
\bibitem [{\citenamefont {Hoang}(2017)}]{Hoang2017PRMoxides}%
  \BibitemOpen
  \bibfield  {author} {\bibinfo {author} {\bibfnamefont {K.}~\bibnamefont
  {Hoang}},\ }\bibfield  {title} {\enquote {\bibinfo {title} {First-principles
  theory of doping in layered oxide electrode materials},}\ }\href@noop {}
  {\bibfield  {journal} {\bibinfo  {journal} {Phys. Rev. Mater.}\ }\textbf
  {\bibinfo {volume} {1}},\ \bibinfo {pages} {in press} (\bibinfo {year}
  {2017})}\BibitemShut {NoStop}%
\bibitem [{\citenamefont {Hoang}\ and\ \citenamefont
  {Johannes}(2011)}]{Hoang2011CM}%
  \BibitemOpen
  \bibfield  {author} {\bibinfo {author} {\bibfnamefont {K.}~\bibnamefont
  {Hoang}}\ and\ \bibinfo {author} {\bibfnamefont {M.}~\bibnamefont
  {Johannes}},\ }\bibfield  {title} {\enquote {\bibinfo {title} {{Tailoring
  Native Defects in \ce{LiFePO4}: Insights from First-Principles
  Calculations}},}\ }\href {\doibase 10.1021/cm200725j} {\bibfield  {journal}
  {\bibinfo  {journal} {Chem. Mater.}\ }\textbf {\bibinfo {volume} {23}},\
  \bibinfo {pages} {3003--3013} (\bibinfo {year} {2011})}\BibitemShut {NoStop}%
\bibitem [{\citenamefont {Lu}\ \emph {et~al.}(2002)\citenamefont {Lu},
  \citenamefont {Beaulieu}, \citenamefont {Donaberger}, \citenamefont
  {Thomas},\ and\ \citenamefont {Dahn}}]{Lu2002JES}%
  \BibitemOpen
  \bibfield  {author} {\bibinfo {author} {\bibfnamefont {Z.}~\bibnamefont
  {Lu}}, \bibinfo {author} {\bibfnamefont {L.~Y.}\ \bibnamefont {Beaulieu}},
  \bibinfo {author} {\bibfnamefont {R.~A.}\ \bibnamefont {Donaberger}},
  \bibinfo {author} {\bibfnamefont {C.~L.}\ \bibnamefont {Thomas}}, \ and\
  \bibinfo {author} {\bibfnamefont {J.~R.}\ \bibnamefont {Dahn}},\ }\bibfield
  {title} {\enquote {\bibinfo {title} {{Synthesis, Structure, and
  Electrochemical Behavior of Li[Ni$_x$Li$_{1/3-2x/3}$Mn$_{2/3-x/3}$]O$_2$}},}\
  }\href {\doibase 10.1149/1.1471541} {\bibfield  {journal} {\bibinfo
  {journal} {J. Electrochem. Soc.}\ }\textbf {\bibinfo {volume} {149}},\
  \bibinfo {pages} {A778--A791} (\bibinfo {year} {2002})}\BibitemShut {NoStop}%
\bibitem [{\citenamefont {Koga}\ \emph {et~al.}(2014)\citenamefont {Koga},
  \citenamefont {Croguennec}, \citenamefont {M\'{e}n\'{e}trier}, \citenamefont
  {Mannessiez}, \citenamefont {Weill}, \citenamefont {Delmas},\ and\
  \citenamefont {Belin}}]{Koga2014JPCC}%
  \BibitemOpen
  \bibfield  {author} {\bibinfo {author} {\bibfnamefont {H.}~\bibnamefont
  {Koga}}, \bibinfo {author} {\bibfnamefont {L.}~\bibnamefont {Croguennec}},
  \bibinfo {author} {\bibfnamefont {M.}~\bibnamefont {M\'{e}n\'{e}trier}},
  \bibinfo {author} {\bibfnamefont {P.}~\bibnamefont {Mannessiez}}, \bibinfo
  {author} {\bibfnamefont {F.}~\bibnamefont {Weill}}, \bibinfo {author}
  {\bibfnamefont {C.}~\bibnamefont {Delmas}}, \ and\ \bibinfo {author}
  {\bibfnamefont {S.}~\bibnamefont {Belin}},\ }\bibfield  {title} {\enquote
  {\bibinfo {title} {{Operando X-ray Absorption Study of the Redox Processes
  Involved upon Cycling of the Li-Rich Layered Oxide
  Li$_{1.20}$Mn$_{0.54}$Co$_{0.13}$Ni$_{0.13}$O$_2$ in Li Ion Batteries}},}\
  }\href {\doibase 10.1021/jp412197z} {\bibfield  {journal} {\bibinfo
  {journal} {J. Phys. Chem. C}\ }\textbf {\bibinfo {volume} {118}},\ \bibinfo
  {pages} {5700--5709} (\bibinfo {year} {2014})}\BibitemShut {NoStop}%
\bibitem [{\citenamefont {Chevrier}\ \emph {et~al.}(2010)\citenamefont
  {Chevrier}, \citenamefont {Ong}, \citenamefont {Armiento}, \citenamefont
  {Chan},\ and\ \citenamefont {Ceder}}]{Chevrier2010}%
  \BibitemOpen
  \bibfield  {author} {\bibinfo {author} {\bibfnamefont {V.~L.}\ \bibnamefont
  {Chevrier}}, \bibinfo {author} {\bibfnamefont {S.~P.}\ \bibnamefont {Ong}},
  \bibinfo {author} {\bibfnamefont {R.}~\bibnamefont {Armiento}}, \bibinfo
  {author} {\bibfnamefont {M.~K.~Y.}\ \bibnamefont {Chan}}, \ and\ \bibinfo
  {author} {\bibfnamefont {G.}~\bibnamefont {Ceder}},\ }\bibfield  {title}
  {\enquote {\bibinfo {title} {Hybrid density functional calculations of redox
  potentials and formation energies of transition metal compounds},}\ }\href
  {\doibase 10.1103/PhysRevB.82.075122} {\bibfield  {journal} {\bibinfo
  {journal} {Phys. Rev. B}\ }\textbf {\bibinfo {volume} {82}},\ \bibinfo
  {pages} {075122} (\bibinfo {year} {2010})}\BibitemShut {NoStop}%
\end{thebibliography}
%
\end{document}